\documentclass[manuscript]{aastex}
\pdfoutput=1
\usepackage{natbib}
\usepackage{lscape}
\usepackage{wasysym}
\usepackage{txfonts}
\usepackage{graphicx}
\usepackage{hyperref}
\bibpunct{(}{)}{;}{a}{}{,}
\usepackage{longtable}

\usepackage{afterpage}

\def\ms{\hbox{\,m\,s$^{-1}$}}         
\def\m2s2{\hbox{\,m$^{2}$\,s$^{-2}$}} 

\newcommand{\titleast}{\ast}
\newcommand{\titlestar}{\star}

\shorttitle{The Kepler-10 planetary system revisited by HARPS-N}
\shortauthors{X. Dumusque and the HARPS-N team}

\begin{document}

\title{
The Kepler-10 planetary system revisited by HARPS-N:\\
A hot rocky world and a solid Neptune-mass planet.\altaffilmark{\titleast}}

\author{Xavier Dumusque\altaffilmark{1}\altaffilmark{\titlestar},
		Aldo S. Bonomo\altaffilmark{2},
		Rapha\"elle D. Haywood\altaffilmark{3}, 
		Luca Malavolta\altaffilmark{4,5},
		Damien S\'egransan\altaffilmark{6},
		Lars A. Buchhave\altaffilmark{1,7},
		Andrew Collier Cameron\altaffilmark{3}, 
		David W. Latham\altaffilmark{1},
		Emilio Molinari\altaffilmark{8,9},
		Francesco Pepe\altaffilmark{6},
		St\'ephane Udry\altaffilmark{6},
		David Charbonneau\altaffilmark{1}, 
		Rosario Cosentino\altaffilmark{8}, 
		Courtney D. Dressing\altaffilmark{1}, 
		Pedro Figueira\altaffilmark{10}, 
		Aldo F. M. Fiorenzano\altaffilmark{8}, 
		Sara Gettel\altaffilmark{1}, 
		Avet Harutyunyan\altaffilmark{8}, 
		Keith Horne\altaffilmark{3}, 
		Mercedes Lopez-Morales\altaffilmark{1}, 
		Christophe Lovis\altaffilmark{6},  
		Michel Mayor\altaffilmark{6}, 
		Giusi Micela\altaffilmark{11}, 
		Fatemeh Motalebi\altaffilmark{6}, 
		Valerio Nascimbeni\altaffilmark{5}, 
		David F. Phillips\altaffilmark{1}, 
		Giampaolo Piotto\altaffilmark{4,5}, 
		Don Pollacco\altaffilmark{12}, 
		Didier Queloz\altaffilmark{6,13}, 
		Ken Rice\altaffilmark{14}, 
		Dimitar Sasselov\altaffilmark{1},  
		Alessandro Sozzetti\altaffilmark{2}, 
		Andrew Szentgyorgyi\altaffilmark{1}, 
		Chris Watson\altaffilmark{15}}

\altaffiltext{1}{Harvard-Smithsonian Center for Astrophysics, 60 Garden Street, Cambridge, Massachusetts 02138, USA}
\altaffiltext{2}{INAF - Osservatorio Astrofisico di Torino, via Osservatorio 20, 10025 Pino Torinese, Italy}
\altaffiltext{3}{SUPA, School of Physics \& Astronomy, University of St. Andrews, North Haugh, St. Andrews Fife,
KY16 9SS, UK}
\altaffiltext{4}{Dipartimento di Fisica e Astronomia "Galileo Galilei", Universita'di Padova, Vicolo dell'Osservatorio 3,
35122 Padova, Italy}
\altaffiltext{5}{INAF - Osservatorio Astronomico di Padova, Vicolo dell'Osservatorio 5, 35122 Padova, Italy }
\altaffiltext{6}{Observatoire Astronomique de l'Universit\'e de Gen\`eve, 51 ch. des Maillettes, 1290 Versoix, Switzerland}
\altaffiltext{7}{Centre for Star and Planet Formation, Natural History Museum of Denmark, University of Copenhagen,
DK-1350 Copenhagen, Denmark}
\altaffiltext{8}{INAF - Fundaci—n Galileo Galilei, Rambla JosŽ Ana Fernandez PŽrez 7, 38712 Bre–a Baja, Spain}
\altaffiltext{9}{INAF - IASF Milano, via Bassini 15, 20133, Milano, Italy}
\altaffiltext{10}{Centro de Astrof\`isica, Universidade do Porto, Rua das Estrelas, 4150-762 Porto, Portugal}
\altaffiltext{11}{INAF - Osservatorio Astronomico di Palermo, Piazza del Parlamento 1, 90124 Palermo, Italy }
\altaffiltext{12}{Department of Physics, University of Warwick, Gibbet Hill Road, Coventry CV4 7AL, UK}
\altaffiltext{13}{Cavendish Laboratory, J J Thomson Avenue, Cambridge CB3 0HE, UK}
\altaffiltext{14}{SUPA, Institute for Astronomy, Royal Observatory, University of Edinburgh, Blackford Hill,
Edinburgh EH93HJ, UK}
\altaffiltext{15}{Astrophysics Research Centre, School of Mathematics and Physics, Queens University, Belfast, UK}

\altaffiltext{$\star$}
{Swiss National Science Foundation Fellow; xdumusque@cfa.harvard.edu}

\altaffiltext{$\ast$}
{Based on observations made with the Italian Telescopio Nazionale Galileo (TNG) operated on the island of La Palma by the Fundaci—n Galileo Galilei of the INAF (Istituto Nazionale di Astrofisica) at the Spanish Observatorio del Roque de los Muchachos of the Instituto de Astrofisica de Canarias.}

\begin{abstract}
Kepler-10b was the first rocky planet detected by the \emph{Kepler}
satellite and confirmed with radial velocity follow-up observations
from Keck-HIRES. The mass of the planet was measured with a precision
of around 30\%, which was insufficient to constrain
models of its internal structure and composition in detail. In
addition to Kepler-10b, a second planet transiting the same star with
a period of 45 days was statistically validated, but the radial
velocities were only good enough to set an upper limit of
20\,$M_{\oplus}$ for the mass of Kepler-10c.  To improve the precision
on the mass for planet b, the HARPS-N Collaboration decided to observe
Kepler-10 intensively with the HARPS-N spectrograph on the Telescopio
Nazionale Galileo on La Palma.  In total, 148 high-quality
radial-velocity measurements were obtained over two observing seasons.
These new data allow us to improve the precision of the mass
determination for Kepler-10b to 15\%. With a mass of
3.33 $\pm$ 0.49\,$M_{\oplus}$ and an updated radius of $1.47^{+0.03}_{-0.02}$\,$R_{\oplus}$,
Kepler-10b has a density of 5.8 $\pm$ 0.8\,g\,cm$^{-3}$, very close to
the value predicted by models with the same internal structure and
composition as the Earth.  We were also able to
determine a mass for the 45-day period planet Kepler-10c, with an even better precision of
11\%.  With a mass of 17.2 $\pm$ 1.9\,$M_{\oplus}$ and radius of
$2.35^{+0.09}_{-0.04}$\,$R_{\oplus}$, Kepler-10c has a density of 7.1 $\pm$ 1.0\,g\,cm$^{-3}$.  
Kepler-10c appears to be the first strong evidence of a class of more massive solid planets with longer orbital periods.
\end{abstract}


\section{Introduction} \label{section:0}

Transiting planets enable a rich variety of opportunities to explore
planetary astrophysics, ranging from the determination of bulk
densities to studies of system architectures and atmospheric
compositions, and even climate and weather.  However, for most of
these studies, a critical first step is the determination of a
dynamical mass that confirms and characterizes the planetary nature of
the candidate. This mass, together with the size of the planet
determined from the light curve, yields the bulk density of the
planet.

The importance of determining dynamical masses for small planets,
either by Doppler spectroscopy or transit-time variations (TTVs) or a
photo-dynamical analysis, was brought into focus by the results for
the five small inner planets transiting Kepler-11, ranging in size
from 1.8 to 4.2 times the diameter of the Earth
\citep[][]{Lissauer-2011a}. The masses derived from TTVs eventually
led to densities ranging from 1.7 down to 0.58\,g\,cm$^{-3}$
\citep[][]{Lissauer-2013}, much less than the Earth's bulk density of
5.5 g cm$^{-3}$, presumably due to extended atmospheres of hydrogen
and helium. It became clear that a better understanding of the
occurrence rate for planets similar to the Earth would require density
determinations for a significant population of small planets.

CoRot-7b was the first small planet that looked like it might be dense
enough to be rocky \citep[][]{Queloz-2009,Leger-2009}. However,
determining its dynamical mass was quite challenging, because strong
stellar activity was perturbing the radial velocities
(RV) of the host star.  This inspired work on ways to reduce the
impact of activity-induced RV signals: pre-whitening by the rotation period and
harmonics \citep[][]{Dumusque-2012,Boisse-2011,Queloz-2009},
correlation with activity indicators \citep[][]{Boisse-2009}, velocity
differences using multiple observations during the same night
\citep[][]{Hatzes-2010}, correlation with photometric variations
\citep[][]{Aigrain-2012}, and treating stellar activity as correlated
noise \citep[][Haywood et al. submitted,]{Feroz-2013}.  This work is
finally leading to a consensus that CoRoT-7b has a density consistent
with a rocky planet. Moreover, learning how to correct the CoRoT photometry and
RV measurements for stellar activity is now paying off for other cases
\citep[e.g.][]{Pepe-2013,Howard-2013b,Dumusque-2012}.

Kepler-10b is in many ways a twin of CoRoT-7b, in terms of size and
orbital period, but has the advantage of transiting an old quiet
star. Still, the mass reported in the discovery paper, 4.56 $\pm$
1.3\,$M_{\oplus}$, has a precision not much better than three sigma
\citep[][]{Batalha-2011}, which is well short of the accuracy needed
to distinguish a rocky planet from a water world, nominally 10\%, 
as we argue below. The discovery paper \citep{Batalha-2011}
also reported the detection of a second transiting planet candidate
with a period of 45 days and radius of 2.23\,$R_{\oplus}$ in the
\emph{Kepler} photometry.  However, no orbital motion at this period
was detected in the observed radial velocities, and only an upper
limit of 20\,$M_{\oplus}$ could be placed on the mass.  With more
\emph{Kepler} data and simultaneous \emph{Spitzer Space Telescope}
photometry, \citet{Fressin-2011} reanalyzed the Kepler-10 system and
statistically validated the planetary nature of Kepler-10c with
{\small BLENDER} \citep[][]{Torres-2011}. This result was not a surprise,
because planet candidates in multi-transit systems are much less
likely to be false positives than candidates in single-transit systems
\citep[][]{Latham-2011}.

Kepler-36 has provided the most accurate mass for a confirmed rocky
planet, based on the extreme TTVs observed in the $Kepler$ photometry
for this system, which boasts two transiting planets in orbits close
to a 6:7 resonance \citep[][]{Carter-2012}. This is an example of a
new approach, a photo-dynamical analysis that uses the rich
information available when three or more bodies all eclipse or transit
each other in compact orbits.  However, this is a bizarre system in
which the neighboring planets have densities that differ by a factor
of 8, and it is not clear how the system formed, or how much longer it
will survive. But, photo-dynamical analysis is a powerful path to
detailed masses and radii for those extremely rare systems with mutual
events and continuous high-quality light curves such as those provided
by $Kepler$.

Kepler-78b is the smallest transiting planet with a well-determined
mass (radius of 1.16\,$R_{\oplus}$ and mass of 1.86\,$M_{\oplus}$),
implying a density very similar to that of the Earth, but with large
uncertainties \citep{Pepe-2013,Howard-2013b}.  This recent
success was enabled by the extremely short orbital period of 8.5
hours, implying a molten surface facing the host star and most likely
a very different history of formation compared to the Earth.

Another important recent result is the heroic effort to determine
masses for a few dozen small planet candidates using HIRES
\citep[][]{Marcy-2014}, with positive detections for about two dozen.
Even though the uncertainties on the masses of these objects are
typically one or two sigma, this is still good enough to verify that
they are planets because of the strong constraints provided by the
photometric ephemerides. As a population, many of them are likely to be
rocky, but the results are not good enough to characterize any
individual planet in detail.

The HARPS-N Collaboration established the goal of measuring masses and
radii with an accuracy sufficient to constrain models of the internal
structure and composition of individual small planets, focusing on
those with compact atmospheres.  For example, an accuracy of nominally
10\% or better in mass and 5\% or better in radius is needed to
distinguish rocky planets with iron cores from those with extensive
water content, according to recent models \citep[][]{Zeng-2013}.  

At the start of HARPS-N science operations in August 2012, Kepler-10b looked
like the best target for pursuing this goal. It had already been
confirmed as a rocky planet, with a mass of 4.56\,$M_{\oplus}$, a radius
of 1.42\,$R_{\oplus}$, and a density of
8.8\,g\,cm$^{-3}$ \citep{Batalha-2011}. In addition, the stellar parameters of 
Kepler-10 are well constrained by asteroseismology \citep[][]{Fogtmann-Schulz-2014}, which gives us an unprecedented precision on the radius of Kepler-10b.
The mass determination was based on 40 precise radial velocity measurements obtained with Keck-HIRES, and
the uncertainty in the density of nearly 30\% was dominated by
the uncertainty in the mass.  Over the 2012 and 2013 observing seasons,
we accumulated nearly four times as many radial velocities for
Kepler-10 \citep[with similar precision, see][]{Pepe-2013,Howard-2013b}.
In this paper, we present the analysis of these new data, which allow us to improve
the uncertainty of the
mass for Kepler-10b by about a factor of two, and to determine precisely the mass of Kepler-10c for the first time.

\section{Observations and data reduction}\label{section:01}

We monitored the radial velocity variation of Kepler-10 with the HARPS-N spectrograph installed on the 3.57-m Telescopio Nazionale Galileo (TNG) at the Spanish Observatorio del Roque de los Muchachos, La Palma Island, Spain \citep[][]{Cosentino-2012}. This instrument is an updated version of the original HARPS planet hunter installed on the 3.6-meter telescope at the European Southern Observatory on La Silla, Chile \citep[][]{Mayor-2003}. Just like its older brother, the HARPS-N instrument is an ultra-stable fiber-fed high-resolution (R=115,000) optical echelle spectrograph optimized for the measurement of very precise radial velocities. The use of a more modern monolithic 4k$\times$4k CCD enclosed in a more temperature stable cryostat, and the use of octagonal fibers for a better scrambling of the incoming light fed into the spectrograph should improve the precision of the instrument compared to HARPS. By observing standard stars of known constant radial velocity during the first year of operation, we estimated the RV precision to be of at least 1\,m\,s$^{-1}$ when not limited by photon noise. When observing fainter stars, we expect a radial-velocity precision of 1.2\,m\,s$^{-1}$ to be achieved in a 1-h exposure on a slowly rotating late-G or K-type dwarf with m$_V$ = 12.

Scientific operations began at HARPS-N in August 2012. Over the first two observing seasons, we obtained 157 radial velocity measurements of Kepler-10. Four observations that were obtained during bad weather conditions had very low signal to noise ($<$\,10) and were rejected. During the first year of operation, in 2012, half of the CCD stopped working and the stellar spectra were only recorded on half of the echelle orders. With  only a subset of the lines used to derive the RV using the cross correlation technique, it is not clear if the data taken with half of the chip can be used or not. We therefore decided to reject those five points that were taken before the CCD was replaced. 
This paper therefore presents 148 RV observations of Kepler-10, that are listed in Table \ref{tab:01}. 

\begin{deluxetable}{cccccccc} 
\tabletypesize{\footnotesize} 
\tablewidth{16.5cm} 
\tablecaption{HARPS-N RV measurements. \label{tab:01}}
\startdata   
			\tableline
			\tableline  BJD - 2400000 & RV & RV error & FWHM & FWHM error & log(R'$_{HK}$) & log(R'$_{HK}$) error & SNR\\	
			 		(d)   & (km\,s$^{-1}$) & (km\,s$^{-1}$) & (km\,s$^{-1}$) & (km\,s$^{-1}$) & (dex) & (dex) & 550nm\\
			\tableline
56072.682384	&	-98.742550	&	0.001760	&	-0.001413	&	0.003948	&	-4.9939	&	0.0212	&	56.20\\
56072.704768	&	-98.742890	&	0.001860	&	-0.010146	&	0.004183	&	-4.9569	&	0.0250	&	53.40\\
56087.575720	&	-98.741520	&	0.002190	&	-0.004506	&	0.004982	&	-4.9606	&	0.0394	&	45.00\\
56087.596901	&	-98.735480	&	0.001980	&	-0.012499	&	0.004465	&	-4.9903	&	0.0334	&	48.50\\
56103.661644	&	-98.740400	&	0.002390	&	-0.013195	&	0.005499	&	-4.9856	&	0.0346	&	38.70\\
\nodata & \nodata & \nodata & \nodata & \nodata & \nodata & \nodata & \nodata \\

		\enddata
		\tablecomments{Table \ref{tab:01} is published in its entirety in the electronic edition of the Astrophysical Journal.
A portion is shown here for guidance regarding its form and content.}
\end{deluxetable}

Kepler-10 has a \emph{Kepler} magnitude $Kp=10.96$ (m$_V$ = 11.16), so exposure times of 30 minutes were required to reach a SNR adequate for high RV precision. Scaling from the expected precision for faint stars given in the first paragraph of this section, 30 minutes of exposure on the slowly rotating G-type dwarf Kepler-10 should yield a precision of 1.12\,m\,s$^{-1}$. Looking at the data, the best precision we could get on a single measurement was 1.16\,m\,s$^{-1}$ for an SNR of 80 at 550\,nm. However, due to not always optimal conditions, the average error on the entire RV set is 1.63\,m\,s$^{-1}$ for an average SNR of 59 at 550\,nm. This precision is two times smaller than the RV semi-amplitude of Kepler-10b \citep[][]{Batalha-2011}, and was expected to be roughly the same as the semi-amplitude of Kepler-10c, assuming that planet has the same mass as the similar-sized planet Kepler-68b \citep[2.31\,$R_{\oplus}$, 8.3\,$M_{\oplus}$][]{Gilliland-2013}.

The data exhibit a small RV offset between the measurements taken with the old and the new CCD mainly because of the different charge transfer efficiency of the detectors. This offset will be considered as a free parameter when fitting the RVs.

\section{Stellar properties} \label{section:1}

Stellar parameters for Kepler-10 have been determined using a combined spectrum with \mbox{SNR} of 750  at 550\,nm, resulting from the co-addition of all the individual spectra with \mbox{SNR} $\geq 30$, after dividing by the blaze function and correcting for the barycentric velocity. 

Equivalent widths (EWs) were measured using a modification of the ARES code \citep[][]{Sousa-2007}, which provides an estimate of the error in the continuum determination and the root mean square of the residuals of the Gaussian fit. These values, together with the full width at half maximum (FWHM) of each line, were used to reject lines with poorly measured EWs. Weak lines with EWs $\leq 5$m\AA\,were excluded from the analysis because of sensitivity to continuum determination. In addition, lines with EW $\geq 110$m\AA\, were also not considered because those lines depart from Gaussian shape.

Atmospheric parameters were determined using the 2013 version of the local thermodynamic equilibrium (LTE) code \mbox{\sc MOOG} \citep[][]{Sneden-1973} and the Kurucz model atmosphere grid\footnote{ Available at http://kurucz.harvard.edu/grids.html} calculated with the new opacity distribution function \citep[\textrm{ODFNEW,}][] {Castelli-2004,Kurucz-1992}. Atmospheric models were interpolated with a software developed by A. McWilliam and I. Ivans and provided by C. Sneden.

We adopted the line list from \citet{Sousa-2011b}, but the oscillator strength values \mbox{log~{\it gf}} were re-determined by inverse analysis of their Solar EWs using an Iron abundance of $\log\,\epsilon(\mathrm{Fe})=  7.52$ (and the other photospheric parameters unchanged), to be consistent with the Solar value used in the calculation of the model atmosphere grid and by \mbox{\sc MOOG} in the abundance computation.

Atmospheric parameters were derived in a classical way. Effective temperature $T_{{\rm eff}}$ was adjusted until there was no dependance of individual \ion{Fe}{1} abundances with the excitation potential (EP) of the lines. The microturbulence velocity \mbox{$\xi_{\rm t}$} was determined by requiring that the \ion{Fe}{1} abundance be independent of the reduced equivalent widths (REW = EW $/\,\lambda$). Surface gravity \mbox{$\log\,g_{\star}$}  was estimated by imposing the ionization equilibrium, i.\,e. by forcing agreement between the abundances derived from \ion{Fe}{1} and \ion{Fe}{2} lines. We cycled between each parameter iteratively until convergence was reached. 
To make sure that our determination was not influenced by the initial guess of the atmospheric parameters, we tested our results against several different input values, finding no significant difference in the derived atmospheric parameters.  

Internal errors were estimated by considering the dispersion around the mean $\sigma$(\ion{Fe}{1}) of $\log \epsilon$(\ion{Fe}{1}) and its effect on the parameter determination. For $\sigma_{T_{\rm eff}}$ we considered the variation in temperature resulting from a slope equal to the ratio between $\sigma$(\ion{Fe}{1}) and the range in EP, and similarly for $\sigma_{\xi_{\rm t}}$ using the range in REW. 
The dependance of the atmospheric parameters on temperature was evaluated by repeating the analysis with the effective temperature fixed at $T_{\rm eff} \pm\sigma_{T_{\rm eff}}$, following \citet{Cayrel-2004}.

To derive the final atmospheric parameters, we fixed the value of \mbox{$\log\,g_{\star}$} to 4.34, as derived in the recent asteroseismology study of \citet{Fogtmann-Schulz-2014}. This choice was made because the gravity of solar-type stars is known to be better constrained by asteroseismology, and because the value derived from EW analysis, \mbox{$\log\,g_{\star}$} = 4.38 $\pm$ 0.8  is fully in agreement. With this value fixed, the final atmospheric parameters for Kepler-10 are $T_{\rm eff} = 5721 \pm 26$\,K, \mbox{$[{\rm Fe}/{\rm H}]$} $= -0.14 \pm 0.02$, and \mbox{$\xi_{\rm t}$} $=1.10 \pm 0.05$\,km\,s$^{-1}$. This result was obtained using 191 \ion{Fe}{1} and \ion{Fe}{2} lines. The same analysis performed on several individual spectra returned values consistent within the errors, i.\,e. co-addition of spectra did not introduce any systematic effect. For comparison, the analysis on the Sun using the HARPS reflection spectrum of Ganymede provided by Sousa et al. (2007) results in  $T_{\rm eff} = 5781 \pm 22$\,K, \mbox{$\log\,g_{\star}$} $=4.45 \pm 0.06$, \mbox{$\xi_{\rm t}$} $=1.03 \pm 0.04$\,km\,s$^{-1}$, \mbox{$[{\rm Fe}/{\rm H}]$} $= 0.00 \pm 0.02$, which are perfectly consistent with the canonical values for the Sun.

To double-check our derived atmospheric parameters, we also used the SPC code \citep[][]{Buchhave-2012} to derive these parameters with a different method. The final values we found with this technique, fixing \mbox{$\log\,g_{\star}$} to 4.34, are $T_{\rm eff}=5695 \pm 50$\,K, \mbox{$[{\rm Fe}/{\rm H}]$}$=-0.16 \pm 0.08$, and $v\,{\rm sin}\,i=0.6 \pm 0.5$\,km\,s$^{-1}$. These results are consistent with the preceding analysis, which provides confidence in the derived atmospheric parameters. 

The adopted $T_{\rm eff}$ and $[{\rm Fe}/{\rm H}]$ are the averages of the values obtained independently with the two aforementioned methods, i.e. $T_{\rm eff}=5708 \pm 28$~K and $[{\rm Fe}/{\rm H}]=-0.15 \pm 0.04$. The mass, radius, surface gravity, and age of Kepler-10 (see Table \ref{tab:1-0}) were afterwards determined by comparing the asteroseismic stellar density $\rho_\star$ as measured by Fogtmann-Schulz et al. (2014), the $T_{\rm eff}$, and $[{\rm Fe}/{\rm H}]$ with the Yonsei-Yale evolutionary tracks (\citealt{Demarque-2004}; see e.g. \citealt{Sozzetti-2007} and \citealt{Torres-2012}) and by performing the same chi-square minimization as in \citet{Santerne-2011}. 
Uncertainties on stellar parameters were estimated through 5\,000 Monte Carlo simulations by assuming independent gaussian errors on $\rho_\star$, $T_{\rm eff}$, and $[{\rm Fe}/{\rm H}]$. Stellar parameters and their $1-\sigma$ errors are listed in Table~\ref{tab:1-0} and are practically identical to those derived by \citet{Fogtmann-Schulz-2014}, as expected, because our effective temperature and stellar metallicity are consistent within $1-\sigma$ with the values found by these authors. No more iterations to re-determine the atmospheric parameters were required because the derived stellar surface gravity, i.e. $\log\,g_{\star}=4.344 \pm 0.004$, is equal to the previously fixed value.
\begin{deluxetable}{lcc} 
\tabletypesize{\footnotesize} 
\tablewidth{12.5cm} 
\tablecaption{Kepler-10 stellar and atmospheric parameters.\label{tab:1-0}}
\startdata   
			\tableline
			\tableline Parameters & Values & References\\ 
			\tableline
			Mass, $M_{\star}$ ($M_{\odot}$)						&	$ 0.910 \pm 0.021 $			& A \\
			Radius, $R_{\star}$ ($R_{\odot}$)						&	$ 1.065 \pm 0.009$			& A \\
			Stellar density, $\overline{\rho}_{\star}$ (g\,cm$^{-3}$)		&	$1.068 \pm 0.004$			& B \\
			Age (Gyr)											&	$10.6^{+1.5}_{-1.3}$		& A \\
			Distance (pc)										&	173 $\pm$ 27				& C \\
			Luminosity, $L_{\star}$ ($L_{\odot}$)					&	1.004 $\pm$ 0.059			& C \\	
			Absolute V magnitude, $M_V$ (mag)					&	4.746 $\pm$ 0.063			& C\\
			Magnitude, $m_V$ (mag)								&	11.157					& D\\
			Color index $(\mathrm{B-V})$							&	0.622					& D\\
			Effective temperature, $T_{\rm eff}$ (K)					&	5708 $\pm$ 28  			& A\\
			Surface gravity, $\log\,g_{\star}$						&	4.344  $\pm$ 0.004 (fixed)	& A\\
			Microturbulence velocity, \mbox{$\xi_{\rm t}$} (km\,s$^{-1}$)	&	1.07 $\pm$ 0.05 			& A \\
			Metallicity, \mbox{$[{\rm Fe}/{\rm H}]$}					&	$-0.15$ $\pm$ 0.04			& A \\
			Activity index, $<$Log(R'$_{HK}$)$>$					&	$-4.96$					& A \\
			Projected rotation, $v\,{\rm sin}\,i$ (km\,s$^{-1}$)				&	0.6 $\pm$ 0.5\quad - \quad2.04 $\pm$ 0.34& A\\
\enddata
		\tablerefs{A: this work, B: \citet{Fogtmann-Schulz-2014}, C: \citet{Fressin-2011}, D: \citet{Everett-2012}}
		\tablecomments{Two values can be found for the $v\,{\rm sin}\,i$. The first one has been derived with SPC \citep[][]{Buchhave-2012} by fitting synthetic spectra to the observed spectra. This type of analysis is not sensitive to $v\,{\rm sin}\,i$ values below $\sim$2\,km\,s$^{-1}$ because of the limit imposed by the spectrograph resolution. The other value is derived using the full width at half maximum of the cross correlation function (see Section \ref{section:2-1-1}). The error bars represent a $1-\sigma$ uncertainty on the parameters.}
\end{deluxetable}

The center-of-mass velocity of Kepler-10 relative to the solar system barycenter is $-98.7$\,km\,s$^{-1}$. This is an unusually large RV that was already noticed in \citet{Batalha-2011}. Given this barycentric radial velocity, the proper motion of 38\,mas\,yr$^{-1}$ \citep[][]{Zacharias-2009}, and the estimated distance of 173 pc \citep[][]{Fressin-2011}, the probability of the star being a member of the thick disk is 96\% \citep[][]{Soubiran-2005}. This is consistent with the old age of the star, i.e. $10.6^{+1.5}_{-1.3}$\,Gyr. Stars of the thick disc are normally more metal poor than Kepler-10, like for example WASP-21 \citep[][]{Bouchy-2010} with a \mbox{$[{\rm Fe}/{\rm H}]$}$=-0.46$. However the separation in metallicity is not such a clear-cut argument \citep[][ and references therein]{Adibekyan-2013b}.

\section{Analysis of the radial velocity data} \label{section:2}

This section is dedicated to the detection of the orbital motion induced by Kepler-10b and c in the HARPS-N RV data, and the best estimation of their masses. We first discuss the activity of the star and its rotational period, as these parameters can give us important information for the selection of the best model to use when fitting the data. Indeed, significant stellar activity will induce RV variations, and in this case, a suitable model should be considered.

\subsection{Kepler-10 stellar activity and induced radial velocity variations} \label{section:2-0}

\begin{figure}
		\includegraphics[width=8cm,height=6cm]{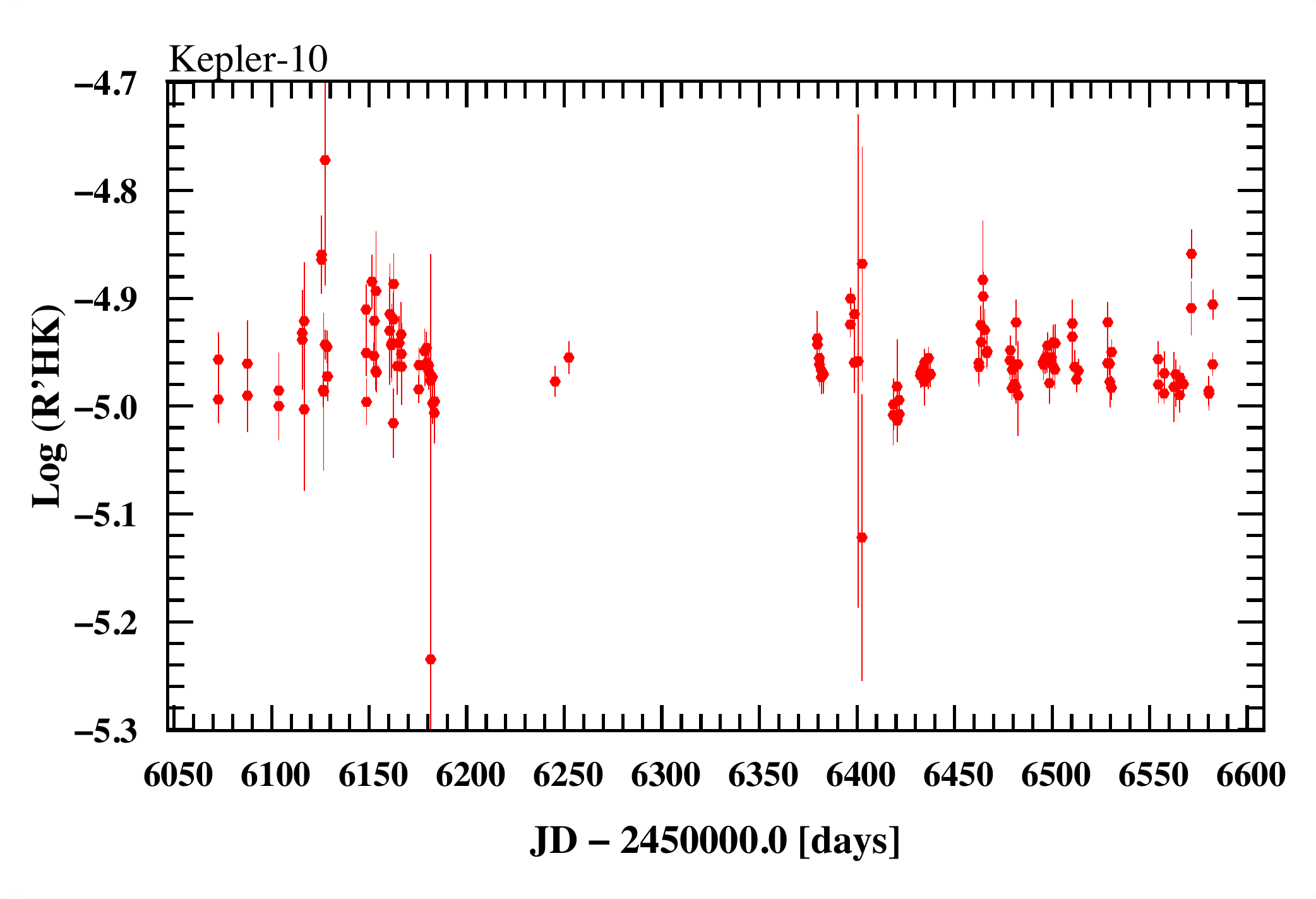}
		\includegraphics[width=8cm,height=5.5cm]{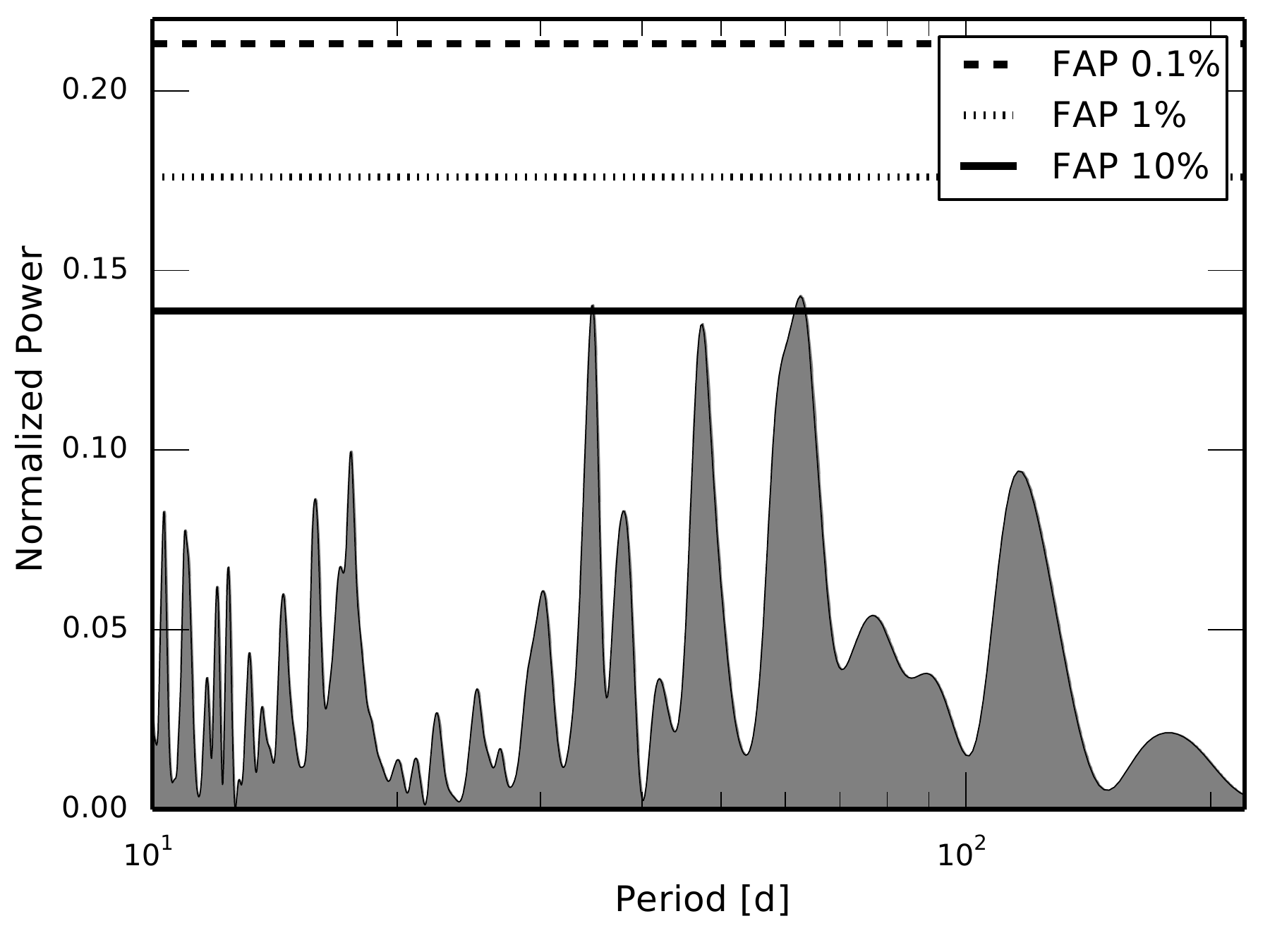}
		\caption{\emph{Left:} Calcium activity index of Kepler-10 as a function of time. The activity level is very low, with an average in log(R'$_{HK}$) equal to $-4.96$, and a standard error of 0.04 dex. \emph{Right: } Periodogram of the calcium activity index. No clear variation in the activity index is detected as no peaks exhibit a FAP lower than 10\%.
		\label{fig:2-0-0}}
\end{figure}

Kepler-10 is a very old main sequence star, estimated to be 10.6 Gyr old, which should imply a low activity level. When measuring the \ion{Ca}{2} HK chromospheric activity index (hereafter calcium activity index) of Kepler-10 \citep[][]{Noyes-1984}, which is the best activity proxy for solar type stars, we find an average value of log(R'$_{HK}$) equal to $-4.96$, with a standard error of 0.04 dex (see Figure \ref{fig:2-0-0}). This average is near the lower end of the solar value, which varies between $-5$ and $-4.8$ along its activity cycle. It is possible that we just observed the star during its minimum activity phase, which represents $\sim$30\% of the solar cycle. However, during the four years of \emph{Kepler} photometry, we do not see any strong variation similar to that induced by typical solar active regions when the Sun is active (see last following paragraph). The study of the activity index points to a star quieter than the Sun, which is compatible with its old age. With this low activity level, we do not expect the RVs to be significantly affected \citep[][]{Dumusque-2011b}.

To estimate the RV variations induced by activity for Kepler-10, we compared the amplitude of the variations seen in the \emph{Kepler} photometry with that expected from active region simulations. Excluding quarters 4 and 17 because they were not detrended by the PDCMAP reduction pipeline \citep[][]{Smith-2012}, and quarter 9 because of instrumental systematics, the maximum peak-to-peak variation in the photometry is less than 500 ppm. Assuming that this variation is due to activity and not to instrumental systematics, and assuming a star with no limb darkening, a 500 ppm variation in photometry would be induced by a dark spot with a filling factor of 0.05\%. According to \citet{Boisse-2012b}, a 1\% dark spot would induce a RV semi-amplitude of 35\,m\,s$^{-1}$ for a $v\,{\rm sin}\,i$ of 3\,km\,s$^{-1}$. Scaling from this value, this spot would induce a 1.75\,m\,s$^{-1}$ semi-amplitude. Considering a more sophisticated model that includes quadratic limb-darkening, convective blueshift suppression in active regions, and a more realistic $v\,{\rm sin}\,i$ of 2\,km\,s$^{-1}$ (see Table \ref{tab:1-0}), this spot would have a filling factor of 0.09\%, and would induce a peak-to-peak RV variation of 1.1\,m\,s$^{-1}$ (Dumusque et al., in prep.). If instead we assume that a plage is inducing this 500 ppm photometric variation, its size would be much larger, 1.4\%, and it would induce a RV peak to-peak variation of 5\,m\,s$^{-1}$. However a plage would also strongly affect the FWHM of the cross correlation function by a peak-to-peak variation of 20\,m\,s$^{-1}$ that is not observed (Dumusque et al., in prep.). In conclusion, the variations observed in the Kepler-10 photometry, if due to activity, would be induced by the presence of spots, and would correspond to a maximum RV peak-to-peak modulation of $\sim$1\,m\,s$^{-1}$, lower than the RV error bars (see Section \ref{section:01}). We therefore do not expect the RVs to be dominated by an activity signal, which is consistent with the low activity index of log(R'$_{HK})=-4.96$ that is measured for the star.

\subsection{Kepler-10 rotational period} \label{section:2-1}

In this section, we estimate the rotational period of Kepler-10 using several different techniques: activity and stellar age estimations, projected rotational velocity, and the \emph{Kepler} photometry.

\subsubsection{Kepler-10 rotational period  estimation with activity index and stellar age} \label{section:2-1-0}

The activity level and rotational period are strongly correlated for main sequence stars. Indeed, stellar activity is generated through the stellar magnetic dynamo, the strength of which appears to scale with rotation velocity \citep[][]{Montesinos-2001,Noyes-1984}. Both stellar activity and rotation are observed to decay with age \citep[][]{Mamajek-2008,Barnes-2007,Pace-2004,Soderblom-1991,Wilson-1963}. In the course of their evolution, solar-type stars lose angular momentum via magnetic braking due to coupling with their stellar wind \citep[][]{Mestel-1968,Weber-1967,Schatzman-1962}.

With an age of 10.6 Gyr, Kepler-10 is a very old main sequence star, which should imply a rotational period longer than the Sun, and therefore a lower activity level.
We saw in the preceding section that the activity level of Kepler-10 is very low, with an average value of log(R'$_{HK})=-4.96$. With this average activity value and a color index $(\mathrm{B-V})= 0.622$ \citep[][]{Everett-2012}, we can estimate a rotational period of 21.9 $\pm$ 3.0 days using the empirical relation of \citet{Mamajek-2008}. The age of 3.7 Gyr given by this empirical relation is much younger than the value derived with asteroseismology. However gyrochronology relations are not well constrained for old main-sequence stars. Given this inconsistency, we can only conclude that the star should be rotating more slowly than a period of 22 days.

Stellar rotation can also be studied by analyzing the periodicity of the variation of the \ion{Ca}{2} HK chromospheric activity index. Active regions coming in and out of view will induce a semi-periodic signal in the activity index and the RVs \citep[][]{Dumusque-2011b}. When looking at a generalized Lomb-Scargle periodogram \citep[GLS,][]{Zechmeister-2009} of the calcium activity index derived with HARPS-N spectra in Figure \ref{fig:2-0-0}, signals with periods greater than 30 days can be found. However these signals all have a false alarm probability (FAP) greater than 10\% implying that there is no significant variation in the activity level, which is generally the case for slow rotators that exhibit a low activity level.

\subsubsection{Kepler-10 rotational period estimation with projected rotational velocity} \label{section:2-1-1}

Another way of estimating the rotational period of the star is using the projected rotational velocity $v\,{\rm sin}\,i$ measured on the stellar spectra, as rotation will broaden spectral lines. As we can see in Table \ref{tab:1-0}, the value of 0.6\,km\,s$^{-1}$ derived by fitting HARPS-N spectra with synthetic spectra is extremely low (using the SPC code). This value is consistent with the one derived in the discovery paper using the SME code \citep[][]{Valenti-1996}. These techniques are not sensitive below $\sim$2\,km\,s$^{-1}$ because of the limited instrumental resolution. Another way of estimating the $v\,{\rm sin}\,i$ is to look at the FWHM of the cross correlation function used to derive the RVs \citep[][]{Boisse-2010,Santos-2002b}. In this case, the instrumental FWHM is 5.9 $\pm$ 0.2\,km\,s$^{-1}$ at $(\mathrm{B-V})= 0.7$, and Kepler-10 has a FWHM of 6.41\,km\,s$^{-1}$. Using the numerical factor of 1.23 from \citet{Hirano-2010} valid for slow rotators, we arrive to $v\,{\rm sin}\,i = \sqrt{\mathrm{FWHM}^2-\mathrm{FWHM_{\mathrm{inst}}}^2}/1.23 = 2.04 \pm 0.34$\,km\,s$^{-1}$. Assuming that the star is seen equator on, which is probable given that we have a multi-planet transiting system \citep[][]{Hirano-2014}, and assuming a radius of 1.065\,$R_{\odot}$, our two $v\,{\rm sin}\,i$ estimates point to slow rotational periods of 90 and 26 days, respectively.

\subsubsection{Kepler-10 rotational period estimation using the \emph{Kepler} photometry} \label{section:2-1-2}

Finally, rotation can also be studied with \emph{Kepler} light curves by searching for flux variations induced by active regions coming in and out of view. As these active regions rotate with the star, a signal with a period similar to the rotation of the star is expected. Depending on the coverage and evolution of active regions, the amplitude of the activity signal can strongly vary from one rotational period to the next because these regions only live for a short amount of time. On the Sun, the lifetime of active regions is about a few rotational periods \citep[e.g.][]{Howard-2000}. 

\emph{Kepler} light curves over time spans of dozens of days are strongly affected by instrumental systematics. As clearly stated in the \emph{Kepler} Data Release 21 Notes, the detrended light curves (PDCSAP) should not be used to look for stellar rotational periods greater than 20 days, as the pipeline removes all signals on longer timescales. Because all the rotation indicators discussed in the preceding sections agree on a rotational period greater than 20 days, we did not attempt to derive the rotational period of Kepler-10 using the \emph{Kepler} photometry.

We used different indicators to infer the rotational period of Kepler-10. None of these indicators can give us a precise value, but they all agree that the star rotates with a period longer than 20 days.

\subsection{Kepler-10b} \label{section:2-2}

\begin{figure}
		\includegraphics[width=9cm]{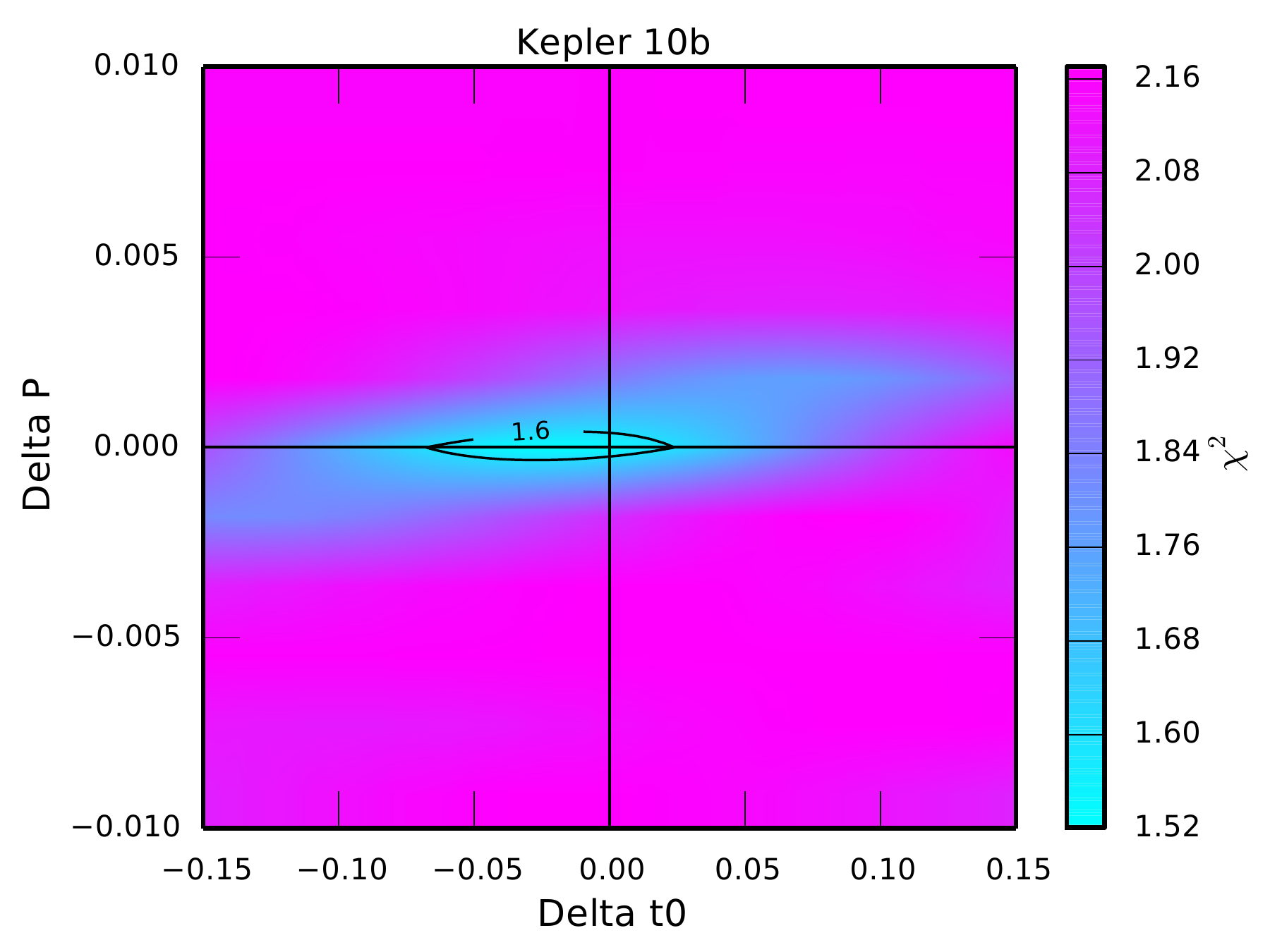}
		\includegraphics[width=9cm]{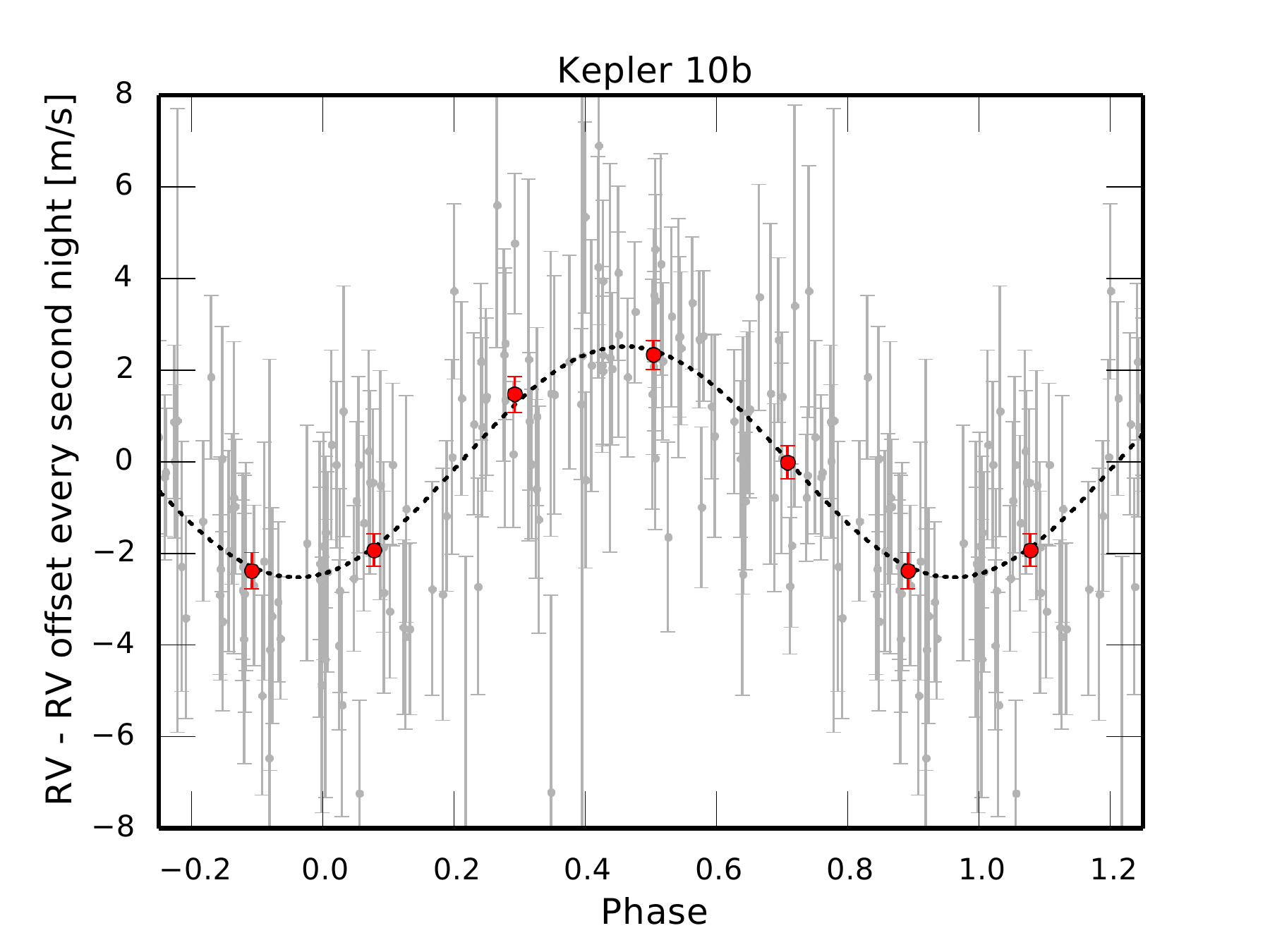}
		\caption{Radial velocity fit of Kepler-10b using a circular orbit model plus RV offsets every two nights. 
		\emph{Left:} Contour plot of the $\chi^2$ surface around the ephemeris for the period and epoch of transit given by the published \emph{Kepler} photometric solution (origin of the axis). The minimum $\chi^2$ fit obtained with the RV measurements matches the photometric solution. 
		\emph{Right:} Phase-folded radial velocity signal of Kepler-10b. The grey points corresponds to all the RV measurements obtained for Kepler-10, the big red dots represent the same data binned in phase with a window of 0.2, and the black dotted line our best fit for Kepler-10b.
			    \label{fig:2}}
\end{figure}

Kepler-10b has an orbital period of 0.84 days, and therefore several measurements per night were obtained to sample efficiently the planetary signal. The other signal expected in the RVs of Kepler-10  is the one coming from Kepler-10c with a 45 day period, if detectable. Considering these signals, the RV variation expected over two days is dominated by Kepler-10b, and as was done for Kepler-78b \citep[][]{Pepe-2013}, we decided to adjust the RV offset every two nights to filter out the signal of Kepler-10c \citep[][]{Hatzes-2010}. Assuming that the eccentricity of Kepler-10b is small \citep[][]{Fogtmann-Schulz-2014}, we fit the RVs with a model composed of a circular orbit plus an RV offset for every two nights. Repeating this optimization over a grid of orbital frequencies and times of mid-transit, we confirm that the RV signal of Kepler-10b is consistent with the photometric transit ephemeris (see Figure \ref{fig:2} \emph{left}).

The best fit for the planetary signal, with a reduced $\chi^2$ of 1.52, converges to a semi-amplitude of $K$ = 2.51$\pm$0.32\,m\,s$^{-1}$. This semi-amplitude, although smaller, is compatible with the discovery solution, $K$ = 3.3$^{+0.8}_{-1}$\,m\,s$^{-1}$. 
The phase-folded RV signal of Kepler-10b, with the RV offset corrected every two nights can be found in Figure \ref{fig:2} \emph{right}.

\subsection{Kepler-10c} \label{section:2-3}

In the preceding section, we analyzed the data without considering Kepler-10c. The signal of this planet, if detectable, was filtered out by adjusting the RV offset every two nights. In this section, we do not utilize this filtering and instead search for significant signals in the generalized Lomb-Scargle periodogram \citep[GLS,][]{Zechmeister-2009} of the raw RVs. 

\begin{figure}
		\includegraphics[width=8cm]{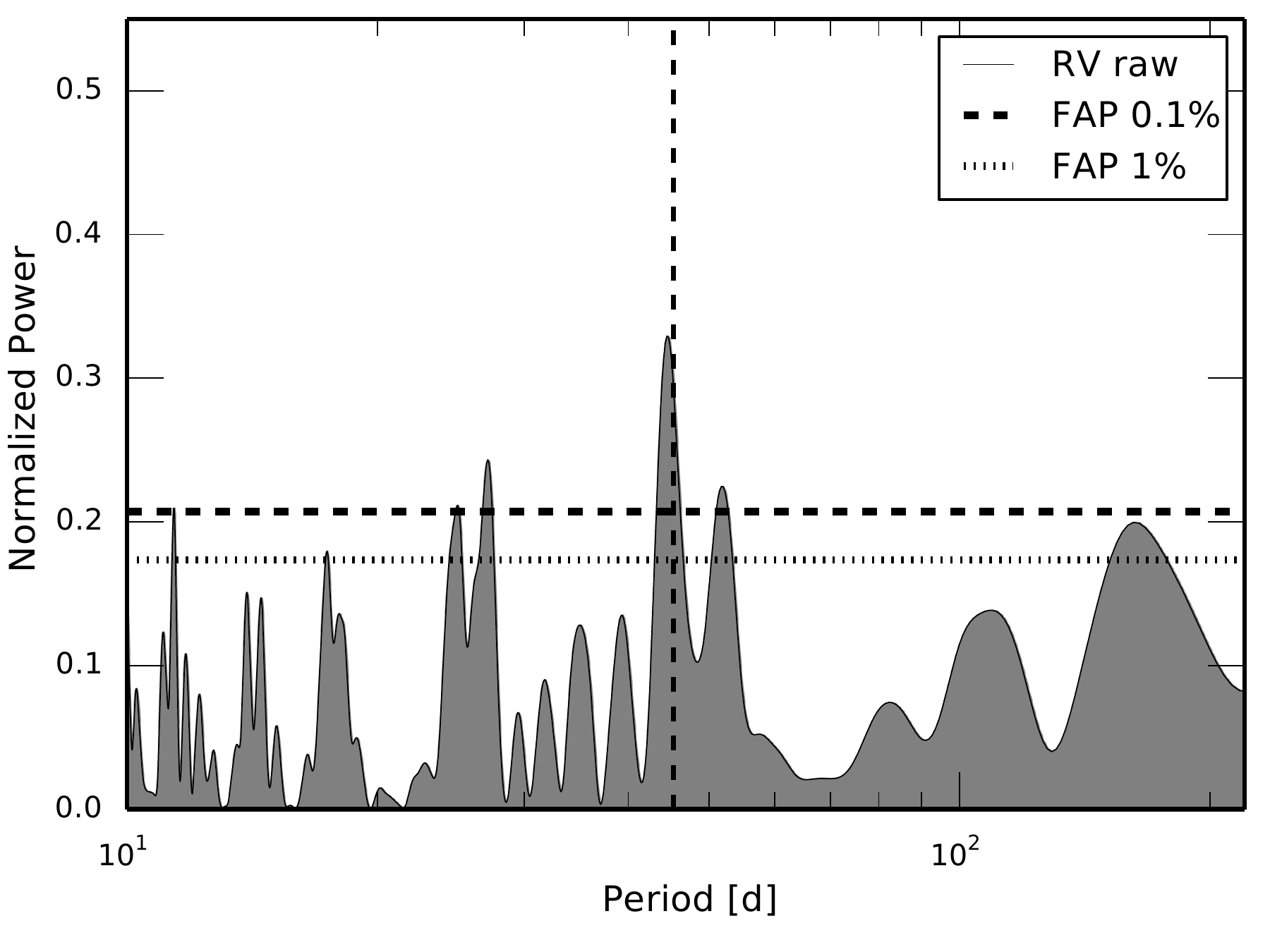}
		\includegraphics[width=8cm]{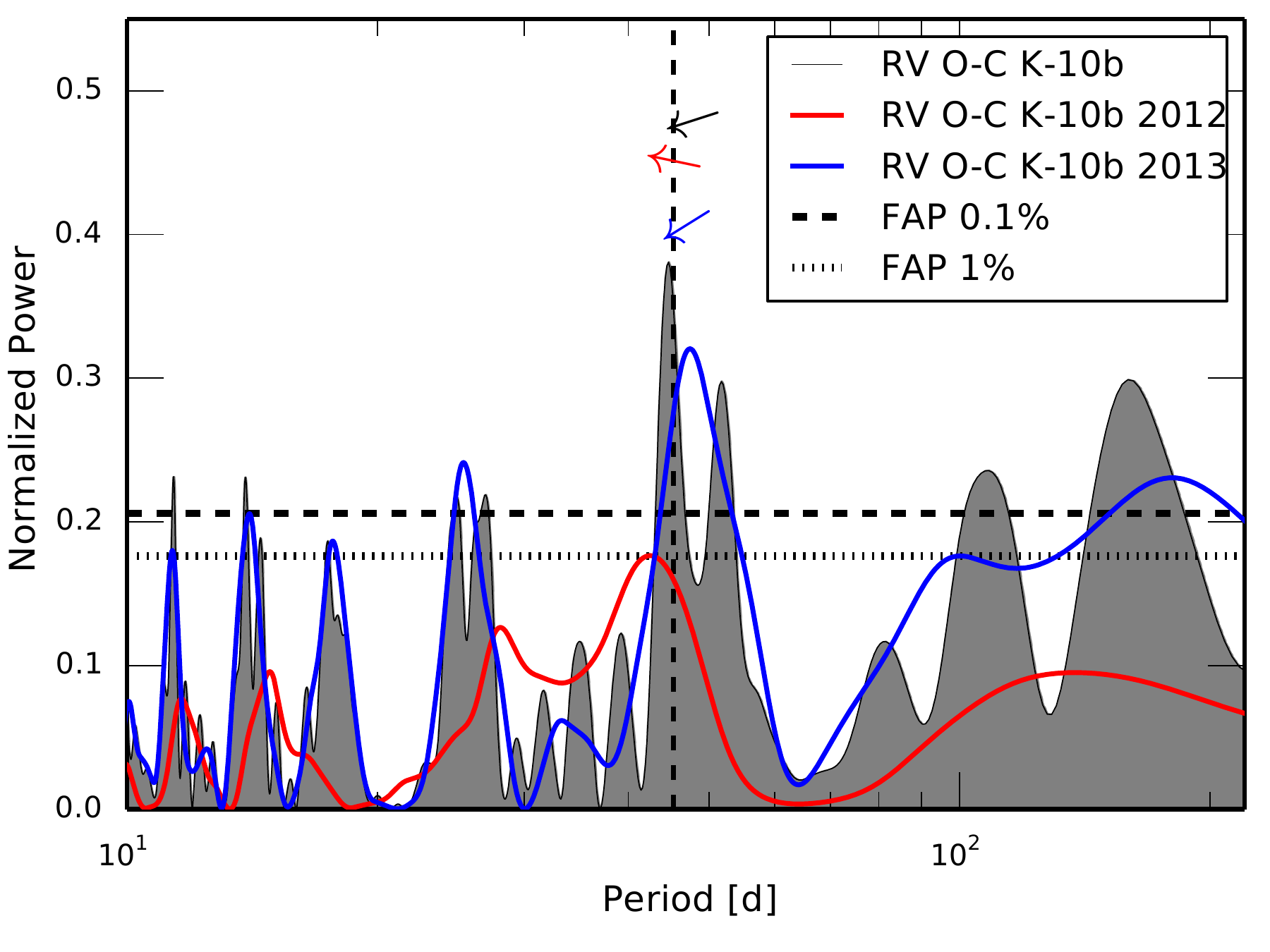}
		\caption{\emph{Left:} GLS periodogram of the raw RVs. The daily aliases of the 45-day peak are seen at $\sim$40 and $\sim$52 days. The photometric period of Kepler-10c is denoted by the vertical dashed line. \emph{Right:} GLS periodogram of the residuals (O-C) after removing the published solution of Kepler-10b. The black shaded periodogram corresponds to the analysis of the entire data set, while the red and blue periodograms correspond to the 2012 and 2013 data, respectively. The small arrows on top of the 45-day peak give the phase of the signal. These arrows can rotate by 360 degrees depending on the phase of the signal. In this case, the phase for the 2012 and 2013 data are compatible between each other and also compatible with the phase found for the entire data set. Therefore this 45-day signal keeps its phase as it is expected for a planet. For each plot, the horizontal lines correspond, from top to bottom, to a FAP level of 0.1\% and 1\%. \label{fig:31}}
\end{figure}

When looking at the GLS periodogram (see Figure \ref{fig:31} \emph{left}), it is clear that a signal at 45 days emerges from the noise, even when the signal of Kepler-10b is not removed. To push the analysis further and check that this 45-day signal corresponds to Kepler-10c and not stellar activity, we removed the signal of Kepler-10b by fitting a circular orbit, fixing the period and the transit time to the published values \citep[][]{Batalha-2011}, and leaving the amplitude as a free parameter. We then compared the GLS periodogram of the RV residuals for the measurements obtained in 2012, the ones obtained in 2013, and for the entire data set. As we can see in Figure \ref{fig:31} \emph{right}, the 45-day signal can be seen in each individual subset. The period of this signal is not well constrained, but it has the same phase from one subset to the other, which is expected for a planetary signal. The semi-amplitudes found, fitting a circular orbit with the published photometric ephemeris of Kepler-10c \citep[][]{Fressin-2011}, are 3.38\,$\pm$\,0.62, 3.74\,$\pm$\,0.44, and 3.11\,$\pm$\,0.35\,m\,s$^{-1}$ for the measurements obtained in 2012, the ones obtained in 2013, and for the entire data set, respectively. The good agreement between these amplitudes, in addition to the fact that the signal keeps the same phase from one year to the next, are strong arguments in favor of the planetary nature of the signal.

\subsection{MCMC analysis of the RVs assuming Gaussian noise with priors from \emph{Kepler} photometry} \label{section:2-3-1}

The preceding section showed that the signal of Kepler-10c is present in the RV data. The next step is to obtain reliable parameters for the mass of this planet\footnote{The orbital inclination of Kepler-10c is 89.7$^{\circ}$ \citep[][]{Fressin-2011}, therefore the minimum mass of the planet is equal to the real mass.}. The \emph{Kepler} photometry exhibits 1124 transits of Kepler-10b and 24 transits of Kepler-10c, which gives us a very-high precision on the period and the transit epoch for both planets. The RV measurements sample $\sim$350 periods of Kepler-10b and $\sim$7 periods of Kepler-10c, with a much sparser sampling. From those numbers, it is clear that the period and the transit time are constrained by the photometry itself and that RV measurements will not help to improve the determination of these parameters. We therefore decided to fit the RV data only, using the strong priors on the planetary periods and times of transit given by the photometry. 
These parameters were estimated by fitting, on all the available \emph{Kepler} short-cadence simple-aperture-photometry data\footnote{http://keplergo.arc.nasa.gov/PyKEprimerLCs.shtml\citep[][]{Jenkins-2010}}  (up to quarter Q17, with a $5-\sigma$ outlier rejection), a model for each transit \citep{Gimenez-2006},
and a straight line to the individual transit times.
The ephemeris values are reported in Table \ref{tab:2-3-2-0} and are consistent
within $1-\sigma$ with those previously derived by \citet{Batalha-2011} 
and later on by \citet{Fogtmann-Schulz-2014}.

The model that is fitted to the RV data includes two Keplerian signals, one for each planet, plus a RV offset $RV_{0}$ to account for a possible offset between the data taken with the old and the new HARPS-N CCD:
\begin{eqnarray}
\Delta RV(t_i) = \gamma + RV_{0}(t_i) + \sum_{j=1}^{j=2} K_j \left[\cos(\theta(t_i,T_{0,j},P_j,e_j) + \omega_j) + e_j\,\cos(\omega_j) \right].
\end{eqnarray}
In this formula, $\gamma$ is the systemic velocity of Kepler-10, and $RV_{0}(t_i)=0$ for $t_i <$ BJD=2456185 and a constant for $t_i >$ BJD=2456185. Each planet $j$ is characterized by its semi-amplitude $K_j$, period $P_j$, time of periastron $T_{0,j}$, eccentricity $e_j$ and argument of periastron $\omega_j$. The function $\theta(t_i,T_{0,j},P_j,e_j)$ is the true anomaly of the planet at time $t_i$. In addition to this model, we also consider that Gaussian noise of stellar origin could affect the RVs. The choice of a Gaussian noise term was made because Kepler-10 does not show any significant sign of activity. Therefore, correlation between measurements due to active regions drifting on the stellar surface is not expected. However, granulation phenomena induced by stellar convection are known to induce correlated noise in the RVs \citep[][]{Dumusque-2011a}. With an effective temperature similar to the Sun, super granulation, which is the most important phenomenon of granulation, should induce a RV rms smaller than 1\,m\,s$^{-1}$ on the timescale of a day. This should perturb the RV signal of Kepler-10b only slightly because the RV semi-amplitude of this planet is three times bigger. In addition, the strategy of observing the target twice per night, over more than 70 nights, allows us to average out this correlated noise. Regarding Kepler-10c, we do not expect any perturbation as the timescale of super granulation and the planetary period are totally different. We concluded that no significant correlated noise should affect the RV measurements of Kepler-10, and we decided to only consider Gaussian noise when fitting the data.

Before starting any fitting procedure on the data, the $T_{0,j}$ values for both planets were shifted close to the average date of the HARPS-N observations to limit error propagation when fitting the periods $P_j$ and epochs $T_{0,j}$. In addition, we used $C_j = \sqrt{e_j}\cos(\omega_j)$ and $S_j = \sqrt{e_j}\sin(\omega_j)$ as free parameters of the fit instead of $e_j$ and $\omega_j$. This modification allows a more efficient exploration of the parameter space in the case of small eccentricities \citep[][]{Ford-2006b}, which is the case for Kepler-10b and c \citep[][]{Fogtmann-Schulz-2014}.
The model is then fitted to the data using a MCMC algorithm that is similar to the one recently applied to the CoRoT-7 system (see Haywood et al. submitted). 
The stellar signal contribution is modeled as a constant jitter term $\sigma_{sj}$ in addition to the RV instrumental noise $\sigma_i$ returned by the HARPS-N pipeline. The following likelihood:
\begin{eqnarray} \label{eq:2-3-1-0}
\mathcal{L} = \frac{1}{\sqrt{2\pi(\sigma_i^2+\sigma_{sj}^2)}}\,\,exp\left[-\frac{\left(RV(t_i)-\Delta_{RV}(t_i)\right)^2}{2 (\sigma_i^2+\sigma_{sj}^2)}\right]
\end{eqnarray}
is used for the MCMC. Uniform priors were set on all the parameters with the constrain that the RV jitter and semi-amplitudes of both Kepler-10b and c must be greater or equal than zero. Gaussian priors for the period and the transit epoch of both planets were imposed based on our previous photometric ephemeris determination (see Table \ref{tab:2-3-2-0}).

\begin{figure}
	\begin{center}
		\includegraphics[width=8cm]{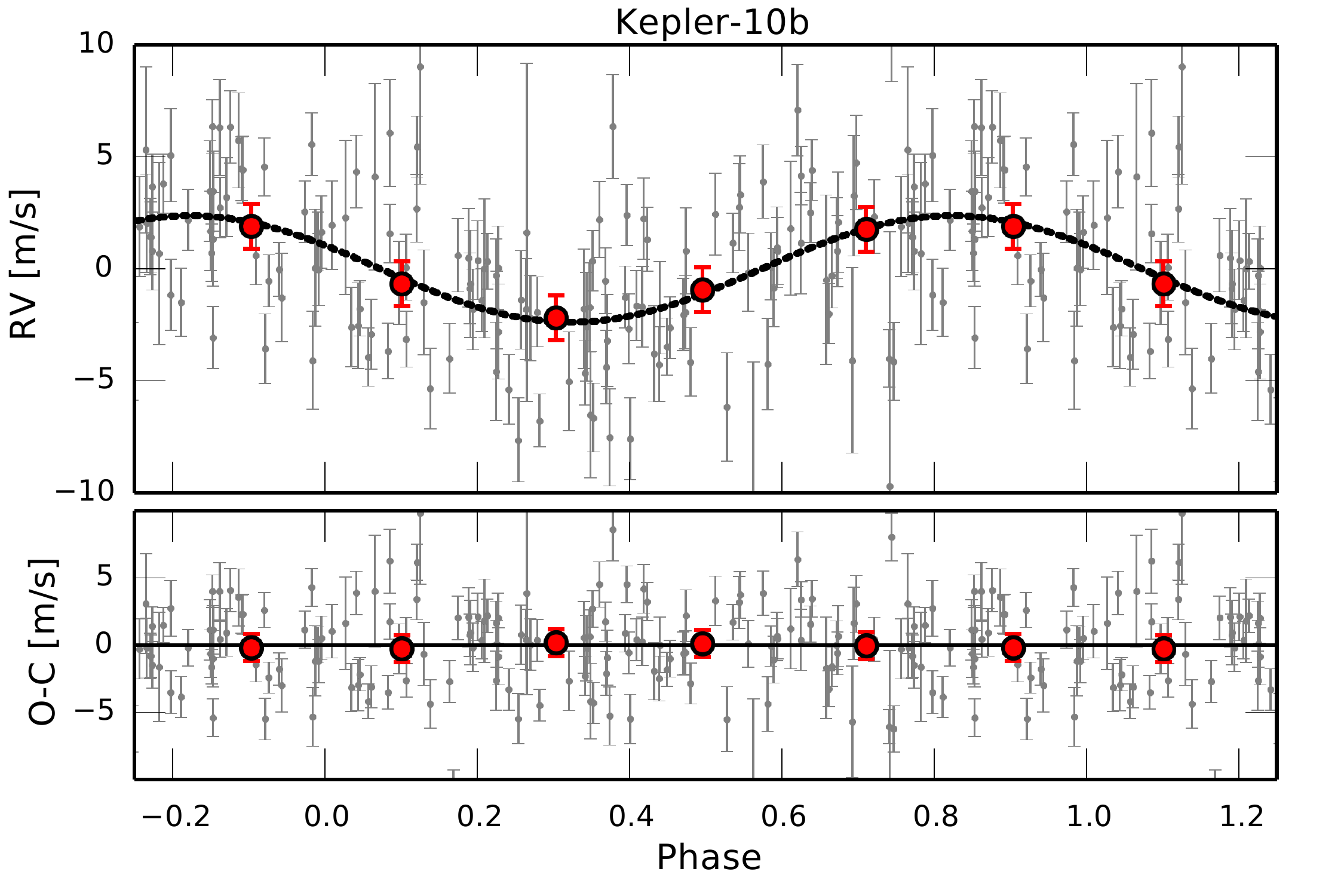}
		\includegraphics[width=8cm]{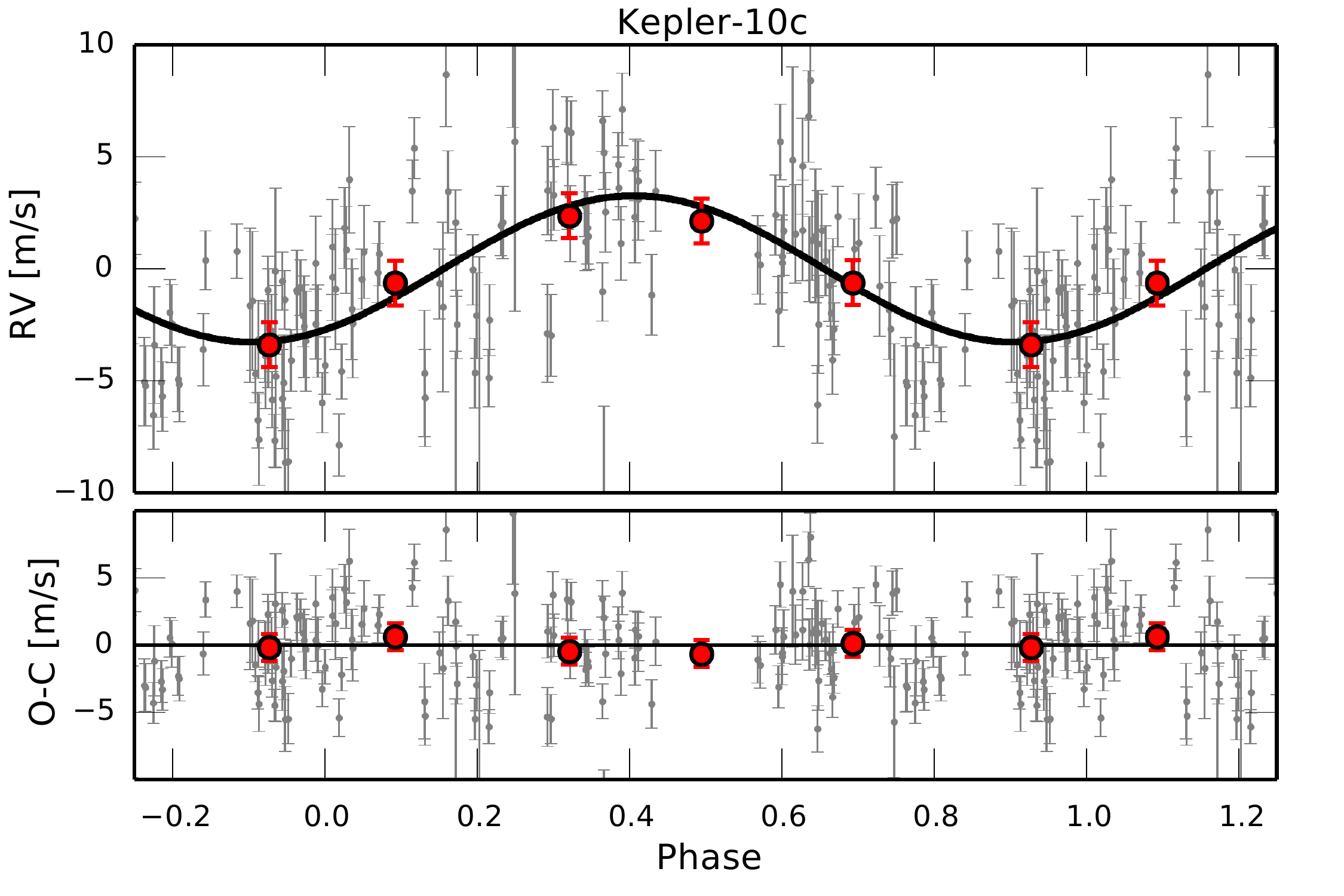}
		\caption{Best phase-folded solution for the fit of Kepler-10b and c on the RVs only, assuming circular orbits and an additional constant stellar jitter. The grey points corresponds to all the RV measurements obtained for Kepler-10, the big red dots represent the same data binned in phase with a window of 0.2, and the black dotted lines our best fit for Kepler-10b and c. \label{fig:2-3-1-0}}
	\end{center}
\end{figure}

\begin{figure}
	\begin{center}
		\includegraphics[width=20cm,angle=90]{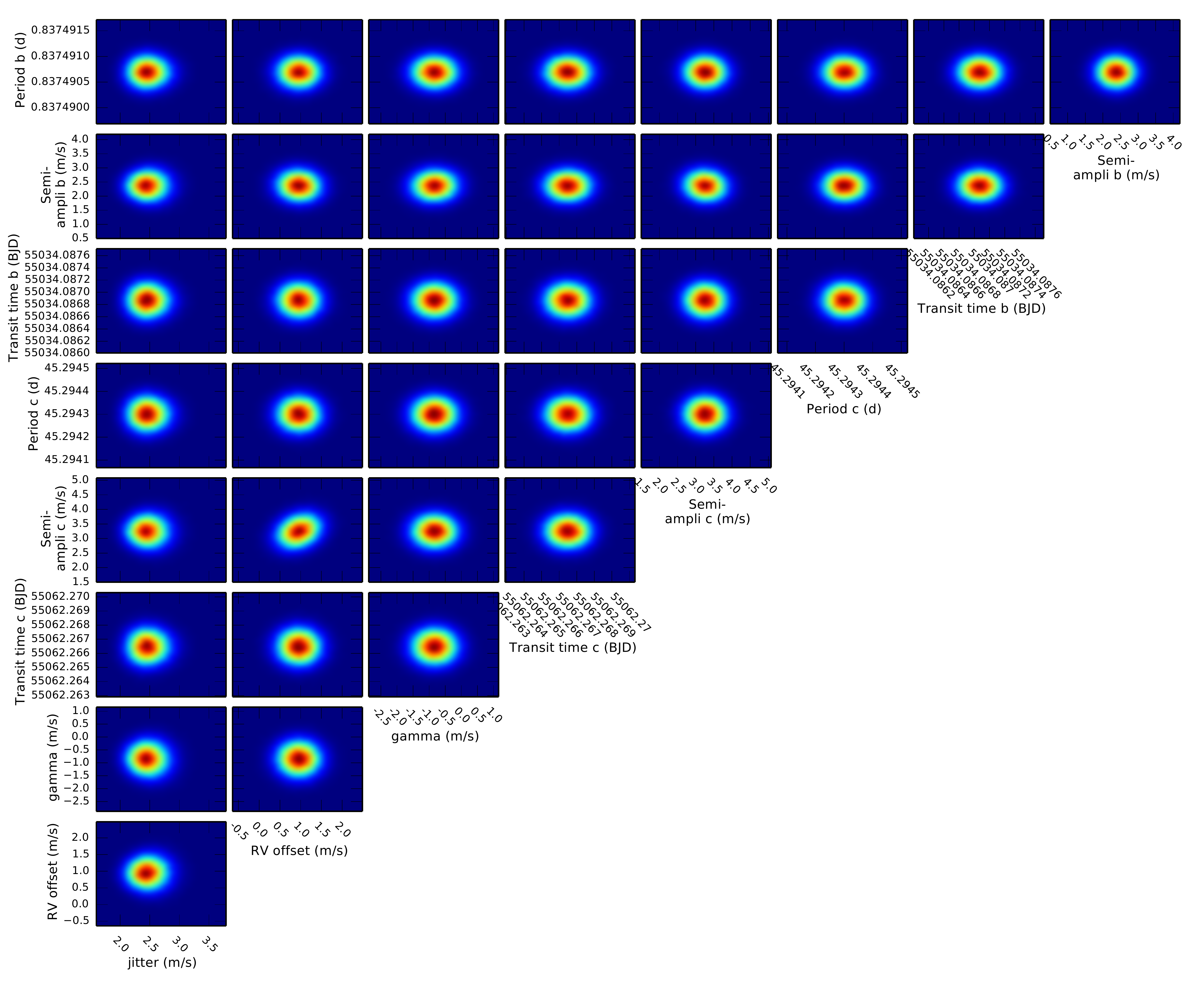}
		\caption{Posterior distributions and correlation between the different MCMC parameters for the model with Gaussian noise and circular orbits. "RV offset" corresponds to the RV offset between the old and the new HARPS-N CCD, while "gamma" is the systemic velocity of Kepler-10.}
		\label{fig:2-3-1-2}
	\end{center}
\end{figure}

Although \citet{Fogtmann-Schulz-2014} find very low eccentricities for both planets, we decided to test a model with free orbital eccentricities (13 parameters) as well as one with eccentricities fixed to zero (9 parameters). In order to assess which model is best, we calculated the marginal likelihood of each model according to the method of \citet{Chib-2001} (see the appendix in Haywood et al. submitted), which uses the posterior distributions of the MCMC chain. 
We obtained a Bayes factor of 10 in favour of the model with fixed zero eccentricities. While this does not constitute strong evidence, according to \citet{Jeffreys-1961}, in favor of circular orbits, it implies that the penalty induced by the extra parameters needed to allow free eccentricities outweighs the increase in likelihood brought by the improvements to the fit. Indeed, when the eccentricities are free to vary, we obtain $e_b < 0.12$ and $e_c < 0.14$ which suggests that both orbits are compatible with circular orbits. The outcome of the circular orbit fit is shown in Table \ref{tab:2-3-2-0}, where the modes of the marginal posteriors with their errors are shown. The phase-folded RVs with the best fit model for Kepler-10b and c can be found in Figure \ref{fig:2-3-1-0}, and the posteriors for all the parameters with their mutual correlation can be found in Figure \ref{fig:2-3-1-2}.

\begin{figure}
	\begin{center}
		\includegraphics[width=7cm]{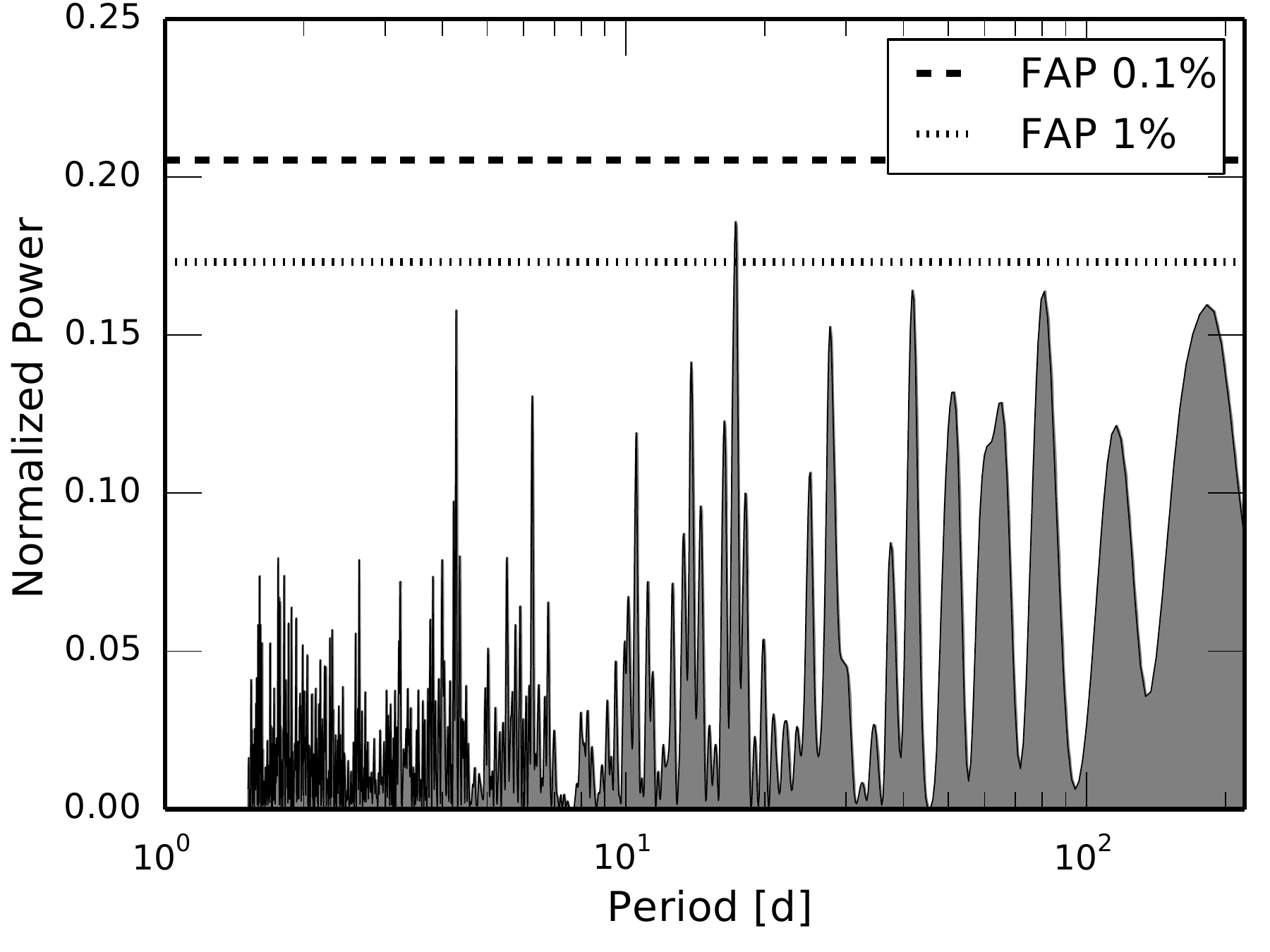}
		\includegraphics[width=7cm]{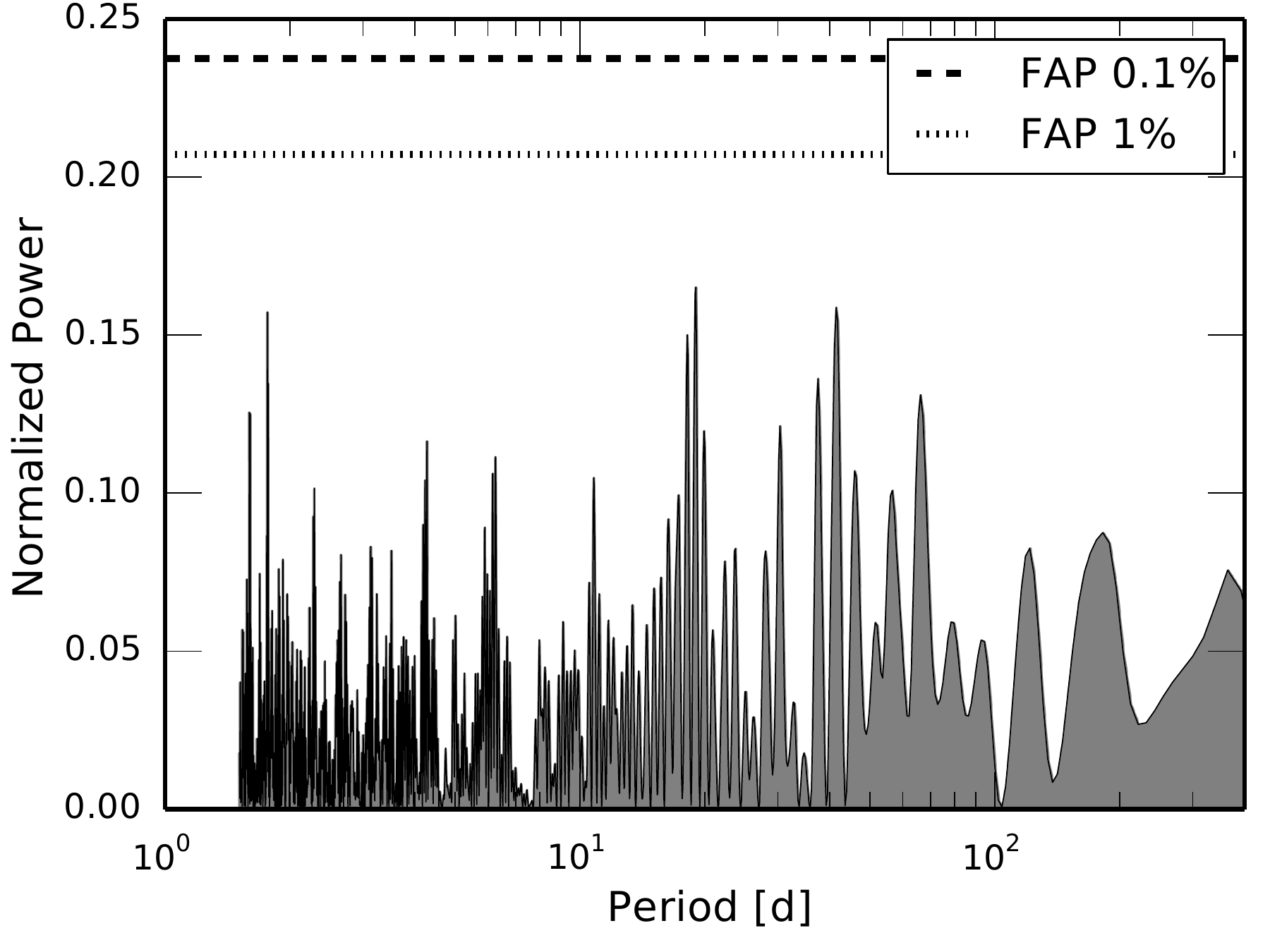}
		\caption{\emph{Left:} GLS periodogram of the RV residuals after fitting to the RVs a circular orbit for both planets and an additional constant stellar jitter. \emph{Right:} Same periodogram but without considering the RV measurements affected by activity, taken between BJD = 2456350 and BJD = 2456450 (31 points in total, equal to 20\% of all the measurements). On both plots, the horizontal lines correspond, from top to bottom, to a FAP level of 0.1\% and 1\%. No significant signal is present in the RV residuals. \label{fig:2-3-1-1}}
			\end{center}
\end{figure}

The RV semi-amplitudes found for Kepler-10b and c are in agreement with our preceding analyses within $1-\sigma$ (see Sections \ref{section:2-2} and \ref{section:2-3}). These semi-amplitudes are estimated with an uncertainty of 14\% and 11\%, respectively. Looking at the GLS periodogram of the RV residuals after removing our best fit solution (see the left plot of Figure \ref{fig:2-3-1-1}), the highest signal remaining in the data is found at $\sim$17 days, with a FAP of 0.5\%. The FAP of this signal is significant enough to try a three-planet model fit. Performing an MCMC with three circular planets converges to a realistic solution. Calculating the Bayes factor between the two and three-planet models results in a highly significant factor of $e^{14}$ in favor of the more complex model. However, looking at the periodogram for each season of observation (see the right plot of Figure \ref{fig:31}), the 17-day signal is only present in the second season. This signal is thus not a coherent signal of constant amplitude spanning both seasons, and therefore cannot be associated to a planet. When removing the RV measurements between BJD = 2456350 and BJD = 2456450 (31 points in total, equal to 20\% of all the measurements), the 17-day signal disappear from the residuals (see the right plot of Figure \ref{fig:2-3-1-1}). These rejected RV measurements seem affected by small activity variations that can be seen when looking at the calcium activity index variation (see left plot of Figure \ref{fig:2-0-0}). The periodogram of the calcium activity index (see right plot of Figure \ref{fig:2-0-0}) shows important, however not significant signals at 34 days and at the first harmonic, i.e 17 days. It is thus possible that the 17-day signal present in the RVs residuals is the first harmonic of the stellar rotation. This 17-day period signal, highly significant when estimating Bayes factor assuming an extra planet, seems to be instead due to short-term activity variations. Even though Bayes factor favors a more complex model including an extra planet, an in-depth analysis is required to characterize the real nature of this extra signal.

\subsection{Combined MCMC analysis of the RVs and \emph{Kepler} photometry considering RV Gaussian noise} \label{section:2-3-2}

In this section, we proceed to a combined RV and photometric fit to determine the orbital and physical parameters of Kepler-10b and c by taking advantage of all the available \emph{Kepler} data.

To derive the Kepler-10 system parameters, 
a Bayesian combined analysis of the \emph{Kepler} short-cadence photometry and 
HARPS-N radial-velocity measurements was performed 
using a Differential Evolution Markov Chain Monte Carlo method
\citep[DE-MCMC,][]{Eastman-2013,Ter-Braak-2006}. 
For this purpose, the epochs of the HARPS-N observations 
were converted from $\rm BJD_{UTC}$ 
into $\rm BJD_{TDB}$ \citep{Eastman-2010}, which is the 
time stamp of \emph{Kepler} data. 
The light curves of Kepler-10b and 10c were phase-folded with 
our new ephemeris (see Table \ref{tab:2-3-2-0}, and discussion in Section \ref{section:2-3-1}) and binned in samples of 4\,s and 30\,s, respectively,
to significantly reduce the computation time 
of our combined analysis.

The HARPS-N radial velocities and the phase-folded and binned light curves
of Kepler-10b and c were simultaneously fitted by considering 
a two-planet model with two Keplerian orbits and two transit models \citep[][]{Gimenez-2006}.
Circular orbits were adopted for both planets based on the
strong constraints from the asteroseismic density determination 
\citep{Fogtmann-Schulz-2014} and our analysis of HARPS-N RV 
measurements in Section \ref{section:2-3-1}.
In total, our model has seventeen free parameters: 
the transit epochs, orbital periods, radial-velocity semi-amplitudes, 
transit durations, orbital inclinations, and planet-to-stellar radius ratios of Kepler-10b
and c, the stellar systemic velocity, a RV offset between the HARPS-N measurements obtained with the old and the new CCD,
an uncorrelated radial-velocity jitter term (e.g., \citealt{Gregory-2005}), and the limb-darkening
coefficients which were fitted by using the triangular sampling method as 
suggested by \citet{Kipping-2013c}: 
$q_{1}=(u_{a}+u_{b})^2$ and $q_{2}=0.5 u_{a} / (u_{a}+u_{b})$, where $u_{\rm a}$ and $u_{\rm b}$ 
are the linear and quadratic coefficients of the limb-darkening quadratic law.

A Gaussian likelihood was maximized in our Bayesian analysis (see Eq. \ref{eq:2-3-1-0}).
Thirty-four DE-MCMC chains, i.e. twice the number of free
parameters, were run simultaneously, 
after being started at different positions in the parameter space 
but reasonably close to the system values.
The jumps of a given chain in the parameter space were determined from two random 
chains, according to the prescriptions given by \citet{Ter-Braak-2006}.
The Metropolis-Hastings algorithm 
was used to accept or reject a proposal step for each chain.
Uniform priors were set on all parameters with the constraints that
the RV jitter and semi-amplitudes of both Kepler-10b
and 10c must be greater or equal than zero, and 
$q_{1}$ and $q_{2}$ are allowed to vary in the 
interval [0, 1] \citep{Kipping-2013c}. Gaussian priors were imposed to
the transit epochs and orbital periods based on our previous 
ephemeris determination (see Table \ref{tab:2-3-2-0}). 

\begin{figure}
	\begin{center}
		\includegraphics[width=7cm]{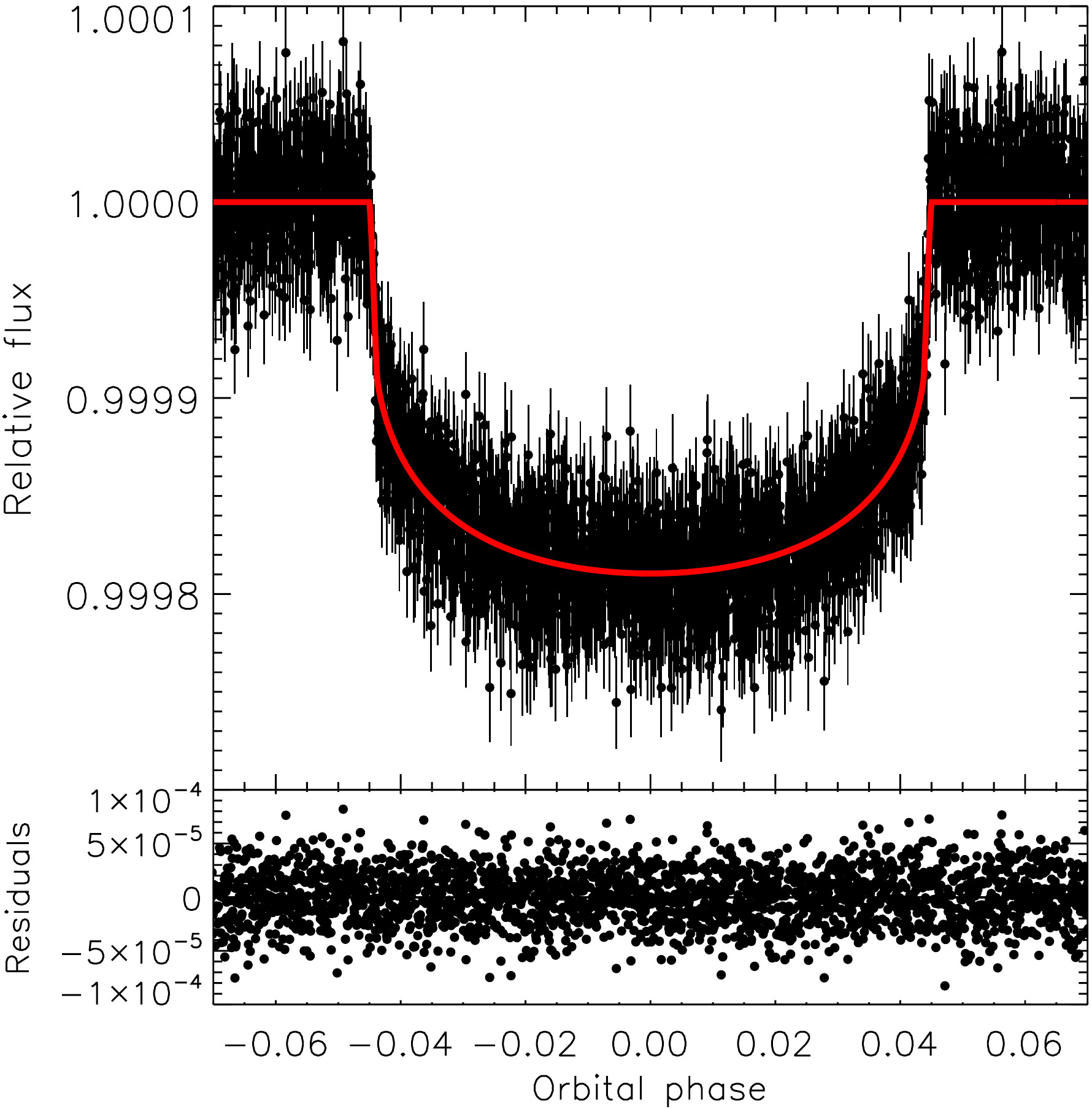}
		\includegraphics[width=7cm]{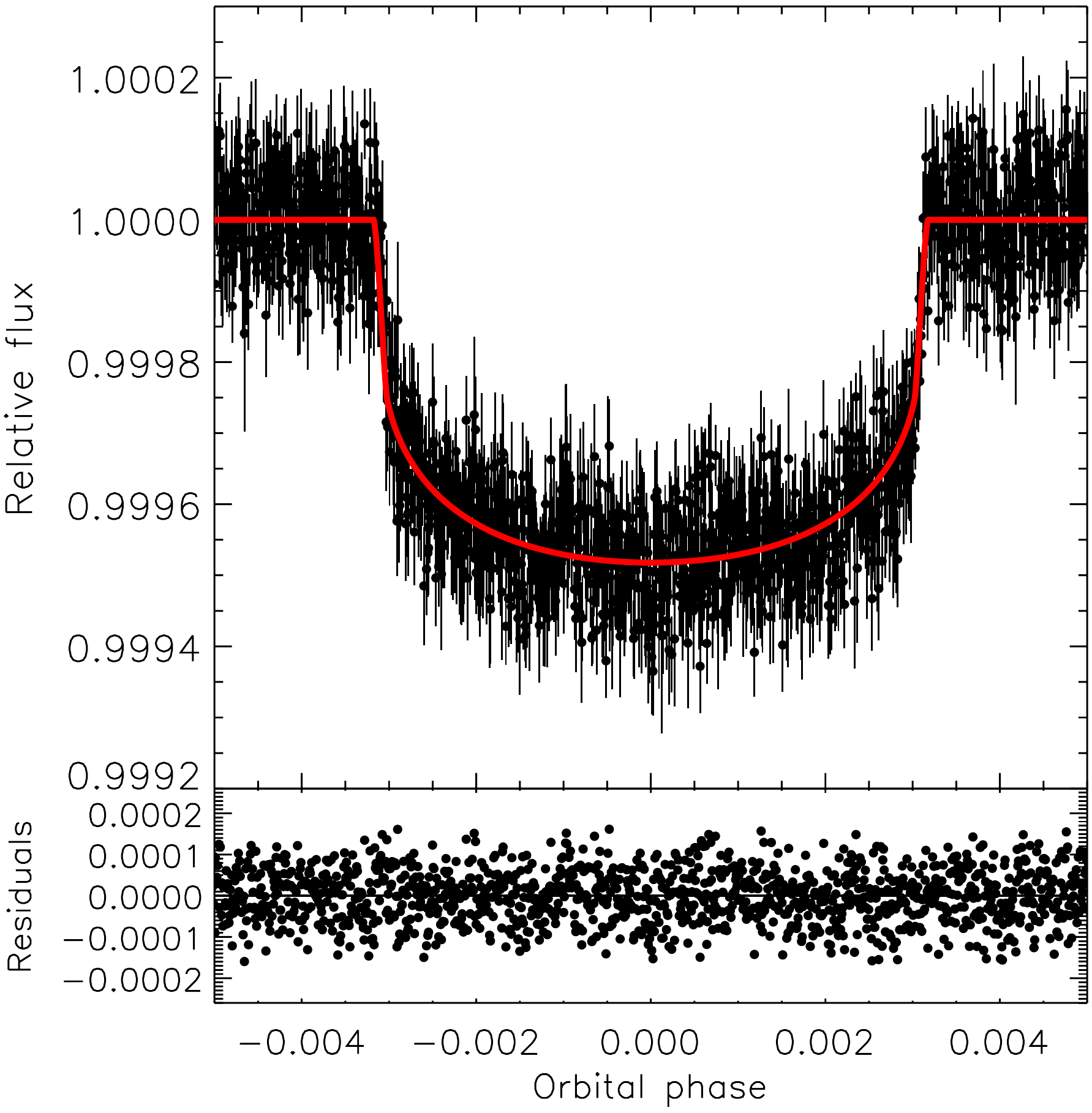}
		\caption{Phase-folded and binned transits of Kepler-10b (\emph{left}) and c (\emph{right}) along with 
				the best-fit models.}
		\label{fig:2-3-1-3}
	\end{center}
\end{figure}

Unlike \citet{Fogtmann-Schulz-2014},
no prior was set on $a/R_\star$ because this prior strongly 
affects the posterior distributions of transit parameters and
yields unrealistic error bars. 
The DE-MCMC analysis was stopped after 
convergence and well mixing of the chains were reached
according to the Gelman-Rubin statistics \citep[$\hat{R} < 1.03$ for all parameters,][]{Gelman-2004}.
Steps belonging to the ``burn-in'' phase were identified 
following \citet{Knutson-2009} and excluded.
Figure \ref{fig:2-3-1-3} shows
the phase-folded and binned transits of Kepler-10b and c along with 
the best-fit models. 
The linear and quadratic limb-darkening coefficients 
$u_{a}=0.40 \pm 0.07$ and $u_{b}=0.27 \pm 0.10$ agree well with values predicted by Kurucz stellar models for 
the \emph{Kepler} bandpass \citep{Sing-2010}, 
i.e. $u_{a}=0.390 \pm 0.006$ and $u_{b}=0.264 \pm 0.004$.
We derived a radius of $\rm 1.47_{-0.02}^{+0.03}~\rm R_{\oplus}$, a mass
of $\rm 3.33 \pm 0.49~M_{\oplus}$, and a density of $\rm 5.8 \pm 0.8~\rm g\;cm^{-3}$
for Kepler-10b and a radius of $\rm 2.35_{-0.04}^{+0.09}~\rm R_{\oplus}$, a mass of $\rm 17.2 \pm 1.9 ~\rm M_{\oplus}$, and a density of
$\rm 7.1 \pm 1.0~\rm g\;cm^{-3}$ for Kepler-10c.

\begin{deluxetable}{l l l} 
\tabletypesize{\scriptsize} 
\tablewidth{16cm} 
\tablecaption{Kepler-10 planetary system parameters.\label{tab:2-3-2-0}} 
\setlength{\tabcolsep}{1cm}
\renewcommand{\arraystretch}{0.7}         
\startdata   
	\tableline
	\tableline
	\emph{Stellar parameters} & RV only & RV and photometry\\
	\tableline             
	Limb-darkening coefficient $u_{a}$ &  \nodata  & $0.40 \pm 0.07$ \\
	Limb-darkening coefficient $u_{b}$ &  \nodata  &  $0.27 \pm 0.10$  \\
	Systemic velocity $\gamma$ (\ms) & $-98739.90 \pm 0.42$ & $-98739.89 \pm 0.43$ \\ 
	RV offset $RV_{0}$ (\ms)  & $1.83 \pm 0.45$ & $1.78 \pm 0.32$ \\  
	Radial-velocity jitter (\ms) & $2.45_{-0.21}^{+0.23}$ & $2.47 \pm 0.23$ \\
	\tableline
	\tableline
	\emph{Kepler-10b }  &  &\\
	\tableline
	\emph{Transit and orbital parameters}  &  \\
	Orbital period $P$ (days) & \multicolumn{2}{c}{$0.8374907 \pm 0.0000002$} \\
	Transit epoch $T_{ \rm 0} (\rm BJD_{TDB}-2454900$) & \multicolumn{2}{c}{134.08687 $\pm$ 0.00018}  \\
	Transit duration $T_{\rm 14}$ (h) & \nodata & $1.8110_{-0.0036}^{+0.0064}$  \\
	Radius ratio $R_{\rm p}/R_{*}$ & \nodata & $0.01261_{-0.00013}^{+0.00026}$   \\
	Inclination $i$ (deg) & \nodata & $84.8_{-3.9}^{+3.2}$  \\
	$a/R_{*}$ & \nodata & $3.46_{-0.28}^{+0.14}$  \\
	Impact parameter $b$ & \nodata  & $0.31 \pm 0.19$  \\
	Orbital eccentricity $e$  &  \multicolumn{2}{c}{0 (fixed)}  \\
	Radial-velocity semi-amplitude $K$ (\ms) & $2.38 \pm 0.34$ & $2.38 \pm 0.35$ \\
	\tableline
	\multicolumn{2}{l}{\emph{Planetary parameters}} \\
	Planet mass $M_{\rm p} ~(\rm M_\oplus)$ & \nodata &  $3.33 \pm 0.49$  \\
	Planet radius $R_{\rm p} ~( \rm R_\oplus)$ & \nodata  &  $1.47_{-0.02}^{+0.03}$  \\
	Planet density $\rho_{\rm p}$ ($\rm g\;cm^{-3}$) & \nodata & $5.8 \pm 0.8$  \\
	Planet surface gravity log\,$g_{\rm p }$ (cgs) & \nodata &  $3.18_{-0.07}^{+0.06}$  \\
	Orbital semi-major axis $a$ (AU) & \nodata & $0.01685 \pm 0.00013$   \\
	Equilibrium temperature $T_{\rm eq}$ (K) & \nodata & $2169_{-44}^{+96}\,^a$ \\
	\tableline  
	\tableline     
	\emph{Kepler-10c }  & & \\
	\tableline
	\emph{Transit and orbital parameters}  &  \\
	Orbital period $P$ [days) & \multicolumn{2}{c}{45.294301 $\pm$ 0.000048} \\
	Transit epoch $T_{ \rm 0} (\rm BJD_{TDB}-2454900$) & \multicolumn{2}{c}{162.26648 $\pm$ 0.00081}  \\
	Transit duration $T_{\rm 14}$ (h) & \nodata & $6.898_{-0.023}^{+0.058}$  \\
	Radius ratio $R_{\rm p}/R_{*}$ & \nodata & $0.02019_{-0.00025}^{+0.00081}$   \\
	Inclination $i$ (deg) & \nodata & $89.59_{-0.43}^{+0.25}$  \\
	$a/R_{*}$ & \nodata & $47.9_{-7.2}^{+2.8}$  \\
	Impact parameter $b$ & \nodata & $0.36_{-0.20}^{+0.25}$  \\
	Orbital eccentricity $e$  &  \multicolumn{2}{c}{0 (fixed)}  \\
	Radial velocity semi-amplitude $K$ (\ms) & $3.25 \pm 0.36$ & $3.26 \pm 0.36$ \\
	\tableline
	\multicolumn{2}{l}{\emph{Planetary parameters}} \\
	Planet mass $M_{\rm p} ~(\rm M_\oplus)$ &  \nodata  &  $17.2 \pm 1.9$  \\
	Planet radius $R_{\rm p} ~( \rm R_\oplus)$ &\nodata  &  $2.35_{-0.04}^{+0.09}$  \\
	Planet density $\rho_{\rm p}$ ($\rm g\;cm^{-3}$) & \nodata &  $7.1 \pm 1.0$  \\
	Planet surface gravity log\,$g_{\rm p }$ (cgs) & \nodata & $3.48_{-0.06}^{+0.05}$  \\
	Orbital semi-major axis $a$ (AU) & \nodata & $0.2410 \pm 0.0019$   \\
	Equilibrium temperature $T_{\rm eq}$ (K) & \nodata & $584_{-17}^{+50}\,^a$ \\
\enddata
\tablerefs{$^{a}$\,black body equilibrium temperature assuming a zero Bond albedo and uniform heat redistribution to the night-side \citep[][]{Rowe-2006}.}
\tablecomments{Kepler-10 planetary system parameters for the circular solution using the RV data only (with priors from photometry), and for the circular solution with the combined photometric and RV analysis. The error bars represent a $1-\sigma$ uncertainty on the parameters.}
\end{deluxetable}

\section{Kepler-10c density} \label{section:3}

\begin{figure}
	\begin{center}
		\includegraphics[width=14cm]{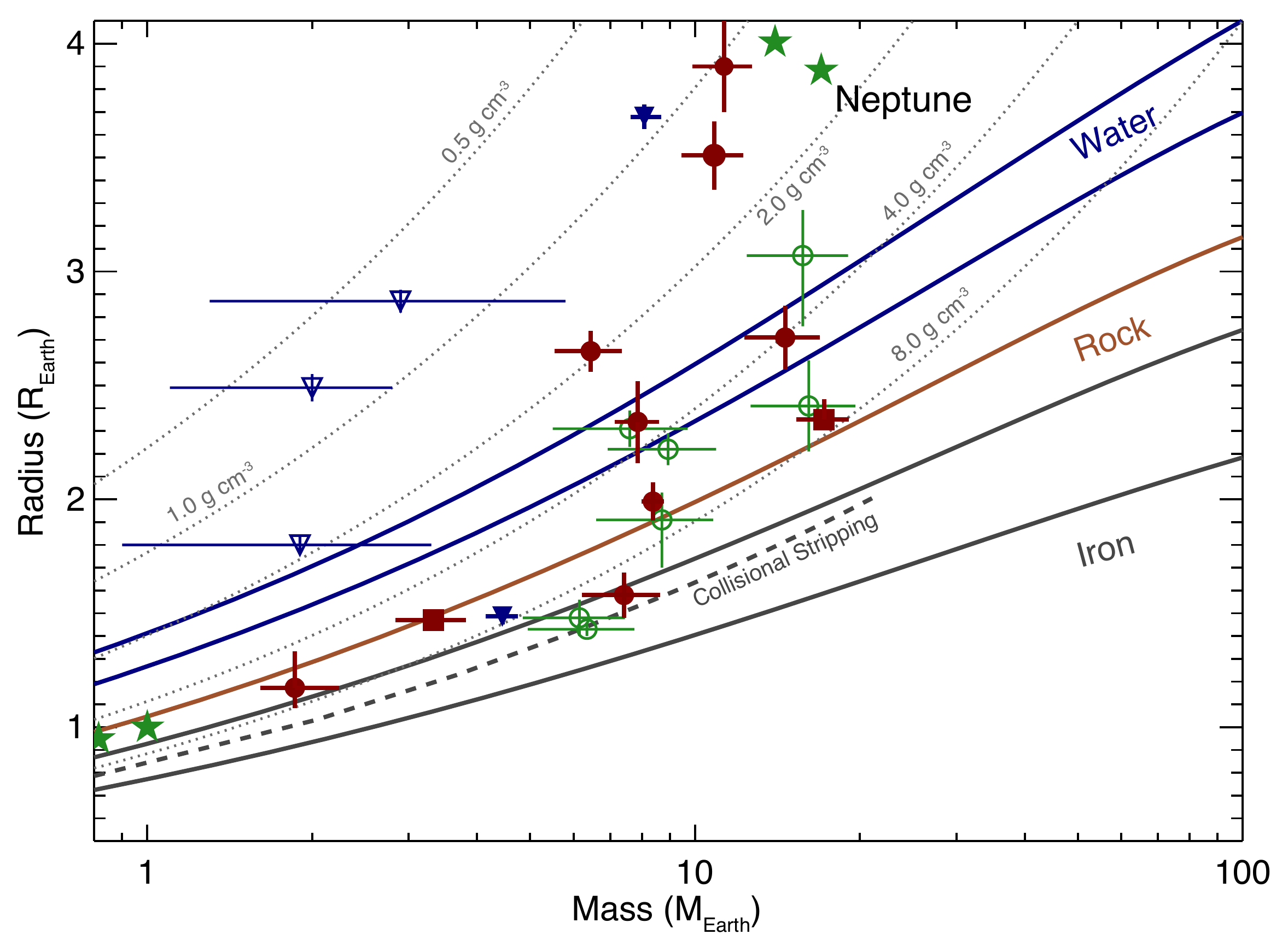}
		\caption{Mass-radius diagram for planets with radius smaller than 4\,$R_{\oplus}$ and that present a mass determination better than 30\%. The only exception are the planets from the Kepler-11 planetary system that we wanted to show here because they have raised the issue that there may be planets with extended envelopes of H and He even at masses less than 5\,$M_{\oplus}$.  Planets with mass determined by the RV technique are represented by circles (except squares to highlight Kepler-10b and c). Planets with mass determined by TTVs are represented by triangles. Filled symbols are used when the precision on the mass is smaller than 20\%, highlighting measurements where an in-depth analysis of the planet composition can be done, while open symbol are used when the precision on the mass is higher than 20\%. The dashed curve is the maximum collisional stripping curve by Marcus et al. (2010), which gives the minimum radii of super-Earths assuming giant impacts during formation. Kepler-10c is the only planet more massive than 10\,$M_{\oplus}$ with such a high density, for which the precision in mass is better than 20\%.}
		\label{fig:6}
	\end{center}
\end{figure}

Understanding the transition from rocky planets to planets defined by a hydrogen-dominated envelope is critical to planet formation theory. Kepler-10c stands out in the mass-radius diagram (Figure \ref{fig:6}), in direct challenge to theory.

To begin with, it is possible to show that Kepler-10c cannot possess a hydrogen-dominated envelope. At 0.24 AU from a solar-luminosity star of very old age, Kepler-10c is too hot ($T_{\rm eq}=584$\,K) to retain an outgassed hydrogen atmosphere, or an accreted hydrogen-helium mixture, as shown by applying \citet{Rogers-2011} calculations for $T_{\rm eq}=500$\,K and the radius of Kepler-10c (all loss timescales are shorter than 1 Gyr). Had Kepler-10c accreted a massive hydrogen-helium envelope 10.6 Gyr ago and still kept some of it, its radius would have to be larger than 3\,$R_{\oplus}$, similar to Kepler-20c (at 15.7$M_{\oplus}$ in Figure \ref{fig:6}). Therein lies the challenge, as Kepler-10c is significantly above the theoretical critical mass of 10-12\,$M_{\oplus}$ enabling envelope accretion \citep[][]{Ikoma-2012, Lopez-2013a}, but has none.

Kepler-10c is best characterized as a solid planet, implying that its bulk composition is dominated by rocks (silicates, judging from the star's composition) and a significant amount of volatile material of high mean molecular weight (i.e. water) of 5\% to 20\%wt. Even if that water is in a separate envelope, which is likely if it is more than 10\% to 15\%wt \citep[][]{Elkins-Tanton-2011}, most of it will be in the form of solid high-pressure ices at age 10.6 Gyr \citep[][]{Zeng-2014}. The precise amount of water in Kepler-10c cannot be constrained further, given the known compositional degeneracy of the mass-radius diagram \citep[][]{Valencia-2007a}.

We note in Figure \ref{fig:6} that Kepler-10c is not alone in this region of the mass-radius diagram, as Kepler-131b \citep[P$_{\rm orb} =$ 16 days, 16.1 $\pm$ 3.5\,$M_{\oplus}$, 2.4 $\pm$ 0.2\,$R_{\oplus}$,][]{Marcy-2014} lies very close to it. However the mass determination of Kepler-131b is not as robust as for Kepler-10c and more data are required to asses if this planet is a twin of Kepler-10c.

Regardless of any water, Kepler-10c is a clear outlier, being many sigmas smaller in size than the densest Neptune-mass exoplanet Kepler-20c. Except for Kepler-131b for which the density still have to be measured more precisely, the exoplanets that appear closest to Kepler-10c in composition, 55 Cnc e and Kepler-20b, are half its mass. They are also both extremely close to their stars and hot, so it is not entirely clear how much the water they contain would contribute to their radii. 

Kepler-10 has a space motion and an age characteristic of the old thick disk population, though with [Fe/H] $= -0.15$, it is not as metal poor as the typical thick disk metallicity of about [Fe/H] $= -0.6$. This implies caution regarding the Fe/Mg and \mbox{Fe/Si} ratios in the proto-planetary disk from which Kepler-10b and 10c formed. The model composition curves on Figure \ref{fig:6} assume solar ratios \citep[][]{Zeng-2013}; the expected variations due to expected enhancements are of 1-2\% in planet radius \citep[][]{Grasset-2009}. Kepler-10b and c might differ in their internal structure in the amount of Fe that is differentiated in their core; for the high mass of Kepler-10c full differentiation is questionable, but given its water content, the oxidation state of Fe is unclear \citep[][]{Zeng-2013}. These uncertainties and the comparison to Kepler-10b imply that Kepler-10c could also have no bulk water, which remains the best composition for Kepler-10b.

\section{Conclusion}  \label{section:3-1}

We revisited the Kepler-10 planetary system using the HARPS-N spectrograph. In total, 148 high-quality RV measurements of the star were obtained, which is nearly four times as many as the Keck-HIRES RV campaign reported in the discovery paper \citep[][]{Batalha-2011}. Using an optimized sampling, we were able to recover the signal of both planets, with a precision on the mass\footnote{It is not surprising that the precision on the mass of Kepler-10b is two times better than the previous value \citep[28\%,][]{Batalha-2011}. With four times as many observations of similar precision \citep[see][]{Pepe-2013,Howard-2013b}, we expect an improvement of a factor of two, everything else being equal.} of 15\% and 11\% for Kepler-10b and c, respectively.

Kepler-10b is a hot rocky world with a mass of $3.33 \pm 0.49$\,$M_{\oplus}$ and a radius of $1.47^{+0.03}_{-0.02}$\,$R_{\oplus}$, which implies a bulk density of 5.8$\pm$0.8\,g\,cm$^{-3}$, slightly higher than the Earth, and an internal structure and composition very similar to Earth.

Kepler-10c is a Neptune mass planet with a density higher than the Earth. Our best estimate of its mass is 17.2$\pm$1.9\,$M_{\oplus}$ for a radius of only $2.35^{+0.09}_{-0.04}$\,$R_{\oplus}$, which yields a density of 7.1$\pm$1.0\,g\,cm$^{-3}$. With these properties, Kepler-10c is best characterized as a solid planet, implying that its bulk composition is dominated by rocks and a significant amount of water, about 5\% to 20\%wt.

Kepler-10c might be the first firm example of a population of solid planets with masses above 10\,$M_{\oplus}$.
From a recent theory of gas accretion at short orbital periods, the critical core mass required to accrete the gas present in the protoplanetary disc onto the planet is $M_{\mathrm{cr}} \sim 2.6\,M{\oplus}\,\left(\frac{\eta}{0.3}\right)^{1/2}\,\left(\frac{P_{\mathrm{orb}}}{1\,\mathrm{day}}\right)^{5/12}$, where $\eta = M_{\mathrm{atm}}/M_{\mathrm{cr}}$ is the fractional mass comprised by the atmosphere \citep[][]{Rafikov-2006}. If this model is correct, it implies that the core mass limit for gas accretion should increase with orbital period. A recent study from Buchhave et al. (2014, in press.), analyzing hundreds of \emph{Kepler} candidates, seems to agree with this theoretical prediction. With a period of 45.29 days, a radius of 2.35\,$R_{\oplus}$, and a density higher than Earth, Kepler-10c would be at the limit of the transition from terrestrial to gaseous planets observed by Buchhave et al. (2014, in press.). Kepler-10c might be the first object confirming that longer period terrestrial planets can be more massive than ones with shorter periods. We note that Kepler-131b \citep[P$_{\rm orb} =$ 16 days, 16.1 $\pm$ 3.5\,$M_{\oplus}$, 2.4 $\pm$ 0.2\,$R_{\oplus}$,][]{Marcy-2014} lies in the same location of the mass-radius diagram as Kepler-10c. However the mass determination of Kepler-131b is not as robust as for Kepler-10c and more data are needed to confirm the high density of this planet. Measuring precisely the mass of several other long-period \emph{Kepler} candidates orbiting bright stars could test this speculation. This experiment has just been started as a new observational program on HARPS-N.

The mass determination of Kepler-10c with HARPS-N shows once more how ground-based RV spectrographs and photometric transit surveys such as \emph{Kepler} can work together to efficiently measure the densities of small planets. Such observational studies are crucial to constrain theoretical models of internal structure and composition of small-radius planets. 

Although Kepler-10c induces a gravitational effect on its host star three times greater than the instrumental precision of HARPS-N, determining its mass to a satisfactory degree of precision required an important RV follow-up campaign; this is because Kepler-10 is a faint star for high-resolution spectrographs and thus the errors are dominated by photon noise. It is essential that future transit searches focus on bright stars in order to allow high-quality RV follow-up needed to measure planet masses with a 10\% precision. The upcoming \emph{Kepler} K2 mission \citep[][]{Howell-2014} should provide soon a few interesting candidates. Then, in a few years, the \small{TESS} satellite \citep[][]{Ricker-2010} should provide hundreds of candidates by performing an all sky survey of short-period transiting planets orbiting stars brighter than V = 12. On the longer term, \small{PLATO} \citep[][]{Rauer-2013} will look for habitable Earth-like planets around those same bright stars.

\acknowledgments
The HARPS-N project was funded by the Prodex Program of the Swiss Space Office (SSO), the Harvard- University Origin of Life Initiative (HUOLI), the Scottish Universities Physics Alliance (SUPA), the University of Geneva, the Smithsonian Astrophysical Observatory (SAO), and the Italian National Astrophysical Institute (INAF), University of St. Andrews, QueenÕs University Belfast and University of Edinburgh. The research leading to these results has received funding from the European Union Seventh Framework Programme (FP7/2007-2013) under Grant Agreement n¡ 313014 (ETAEARTH). 
X. Dumusque would like to thanks the Swiss National Science Foundation (SNSF) for its support through an Early Postdoc.Mobility fellowship.
P. Figueira acknowledges support by  Funda\c{c}\~ao para a Ci\^encia e a Tecnologia (FCT) through the Investigador FCT contract of reference IF/01037/2013 and POPH/FSE (EC) by FEDER funding through the program ''Programa Operacional de Factores de Competitividade - COMPETE''. 
R.D. Haywood acknowledges support from an STFC postgraduate research studentship.
This publication was made possible through the support of a grant from the John Templeton Foundation. The opinions expressed in this publication are those of the authors and do not necessarily reflect the views of the John Templeton Foundation.
This research has made use of the results produced by the PI2S2 Project managed by the Consorzio COMETA, a co-funded project by the Italian Ministero dellÕIstruzione, Universit\`a e Ricerca (MIUR) within the Piano Operativo Nazionale Ricerca Scientifica, Sviluppo Tecnologico, Alta Formazione (PON 2000Ð2006).
We would like to thank A. McWilliam, I. Ivans and C. Sneden for providing us their software that interpolates between atmospheric models.

\bibliographystyle{apj}
\bibliography{dumusque_bibliography}

\end{document}